\documentclass{ecai} 

\usepackage{latexsym}
\usepackage{amssymb}
\usepackage{amsmath}
\usepackage{amsthm}
\usepackage{booktabs}
\usepackage{enumitem}
\usepackage{graphicx}
\usepackage{color}
\usepackage{multirow}
\usepackage{algorithm}
\usepackage{algorithmic}

\newtheorem{definition}{Definition}

\newcommand{\BibTeX}{B\kern-.05em{\sc i\kern-.025em b}\kern-.08em\TeX}

\begin{document}


\begin{frontmatter}


\paperid{3449} 


\title{Learning Macroeconomic Policies through Dynamic Stackelberg Mean-Field Games}



\author[A,B]{\fnms{Qirui}~\snm{Mi}}
\author[A,B]{\fnms{Zhiyu}~\snm{Zhao}}
\author[C]{\fnms{Chengdong}~\snm{Ma}}
\author[A,B]{\fnms{Siyu}~\snm{Xia}}
\author[A]{\fnms{Yan}~\snm{Song}}
\author[D]{\fnms{Mengyue}~\snm{Yang}}
\author[E]{\fnms{Jun}~\snm{Wang}}
\author[A,B,F]{\fnms{Haifeng}~\snm{Zhang}\thanks{Corresponding author. Email: haifeng.zhang@ia.ac.cn.}}

\address[A]{Institute of Automation, Chinese Academy of Sciences, China}
\address[B]{School of Artificial Intelligence, University of Chinese Academy of Sciences, China}
\address[C]{Institute for Artificial Intelligence, Peking University, China}
\address[D]{University of Bristol, UK}
\address[E]{Department of Computer Science, University College London, UK}
\address[F]{Nanjing Artificial Intelligence Research of IA, China}



\begin{abstract}
Macroeconomic outcomes emerge from individuals' decisions, making it essential to model how agents interact with macro policy via consumption, investment, and labor choices. We formulate this as a dynamic Stackelberg game: the government (leader) sets policies, and agents (followers) respond by optimizing their behavior over time. Unlike static models, this dynamic formulation captures temporal dependencies and strategic feedback critical to policy design. However, as the number of agents increases, explicitly simulating all agent–agent and agent–government interactions becomes computationally infeasible. To address this, we propose the \textbf{Dynamic Stackelberg Mean Field Game (DSMFG)} framework, which approximates these complex interactions via agent–population and government–population couplings. This approximation preserves individual-level feedback while ensuring scalability, enabling DSMFG to jointly model three core features of real-world policymaking: \textit{dynamic feedback}, \textit{asymmetry}, and \textit{large scale}. We further introduce \textbf{Stackelberg Mean Field Reinforcement Learning (SMFRL)}, a data-driven algorithm that learns the leader's optimal policies while maintaining personalized responses for individual agents. Empirically, we validate our approach in a large-scale simulated economy, where it scales to 1,000 agents (vs. 100 in prior work) and achieves a $4\times$ GDP gain over classical economic methods and a $19\times$ improvement over the static 2022 U.S. federal income tax policy.
\end{abstract}

\end{frontmatter}


\section{Introduction}

Macroeconomic policy formulation is fundamental to achieving sustainable economic growth~\cite{schneider1988politico,persson1999political}. The effectiveness of these policies depends crucially on the behaviors of micro-level individuals, such as labor supply, consumption, and investment decisions~\cite{sachs1995economic}. Nobel laureate Lucas has emphasized that individuals adapt their decision-making in response to changes in macroeconomic policies~\cite{lucas1976econometric}. Thus, systematically modeling the interactions between individuals and the government is crucial for designing effective macroeconomic policies.

The Stackelberg game naturally captures the asymmetric interactions~\cite{dayanikliMachineLearningMethod2024,huangMeanFieldStackelberg2020} where the government (leader) sets a tax policy, and individuals (followers) adjust their labor supply and consumption in response. Unlike static models, this dynamic formulation accounts for the time-dependent strategic feedback essential for effective policy design~\cite{nashEquilibriumPointsNperson1950,baddeley2019behaviouralmacroeconomicpolicynew,dammann2024stochasticnonzerosumgamecontrolling}. However, scaling this approach to large populations is computationally intractable: with \(N\) agents, there are \(O(N^2)\) pairwise agent-agent interactions plus \(O(N)\) leader–agent interactions.


To address the \textbf{scalability} challenge inherent in macroeconomic policy modeling, we propose the \textbf{Dynamic Stackelberg Mean Field Game (DSMFG)} framework, which unifies \textit{dynamic feedback}, \textit{asymmetry}, and \textit{large scale} into a coherent formulation. Unlike standard SMFG methods—e.g., single-step models~\cite{dayanikliMachineLearningMethod2024} or approaches with fixed dynamics~\cite{bergault2023mean}—DSMFG captures the multi-period feedback loops essential to the co-evolution of policy and individual response. By embedding a multi-step Stackelberg game within a mean-field approximation~\cite{cardaliaguetMeanFieldGame2017,angiuliReinforcementLearningMean2021b}, DSMFG reduces the \(O(N^2)\) complexity of agent-agent interactions to \(O(N)\) agent–population interactions. At each timestep, the government optimizes policy based on the current mean field—i.e., the population’s state–action distribution—while agents adapt to both the policy and the mean field. This iterative structure preserves individual feedback, reduces computational cost, and enables scalable optimization in complex macroeconomic environments.

To learn optimal policies under DSMFG, we propose the \textbf{Stackelberg Mean Field Reinforcement Learning (SMFRL)} algorithm. SMFRL introduces a Stackelberg Mean Field Q-function that enables the leader to evaluate its interactions with the aggregate population. Another Q-function for followers evaluates their interactions with both the leader and the population. Followers share a common policy that takes heterogeneous features as input, enabling personalized actions while maintaining population-level consistency. The central Q and shared policy are designed to ensure scalability in large populations. Moreover, SMFRL employs alternating updates between the leader’s and followers’ policies to ensure a stable training.

We empirically evaluate DSMFG and SMFRL in a large-scale macroeconomic simulation environment, TaxAI, which models dynamic interactions between the government and large scale agents. Comparing against static policies (e.g., the 2022 U.S.\ federal tax) and dynamic rule-based methods (e.g., the Saez tax), DSMFG yields substantially better outcomes—achieving a \(4\times\) gain in per capita GDP over the Saez tax and a \(19\times\) improvement over the 2022 U.S.\ baseline. Unlike static or independent-agent baselines, DSMFG ensures sustainability and rapidly stabilizes income and consumption after shocks. Ablation studies confirm the necessity of both Stackelberg hierarchy and mean-field approximation: removing either leads to lower welfare, instability, and poor convergence. Despite sharing a policy, followers maintain robust performance under population heterogeneity, enabling scalable training and consistently high utility. These results demonstrate DSMFG’s effectiveness in solving dynamic, large-scale government–agent problems within a controlled, reproducible simulation framework, extending beyond the scope of traditional approaches.

In summary, Our key contributions are:
\begin{figure*}[h]
    \centering
    \includegraphics[width=0.9\linewidth]{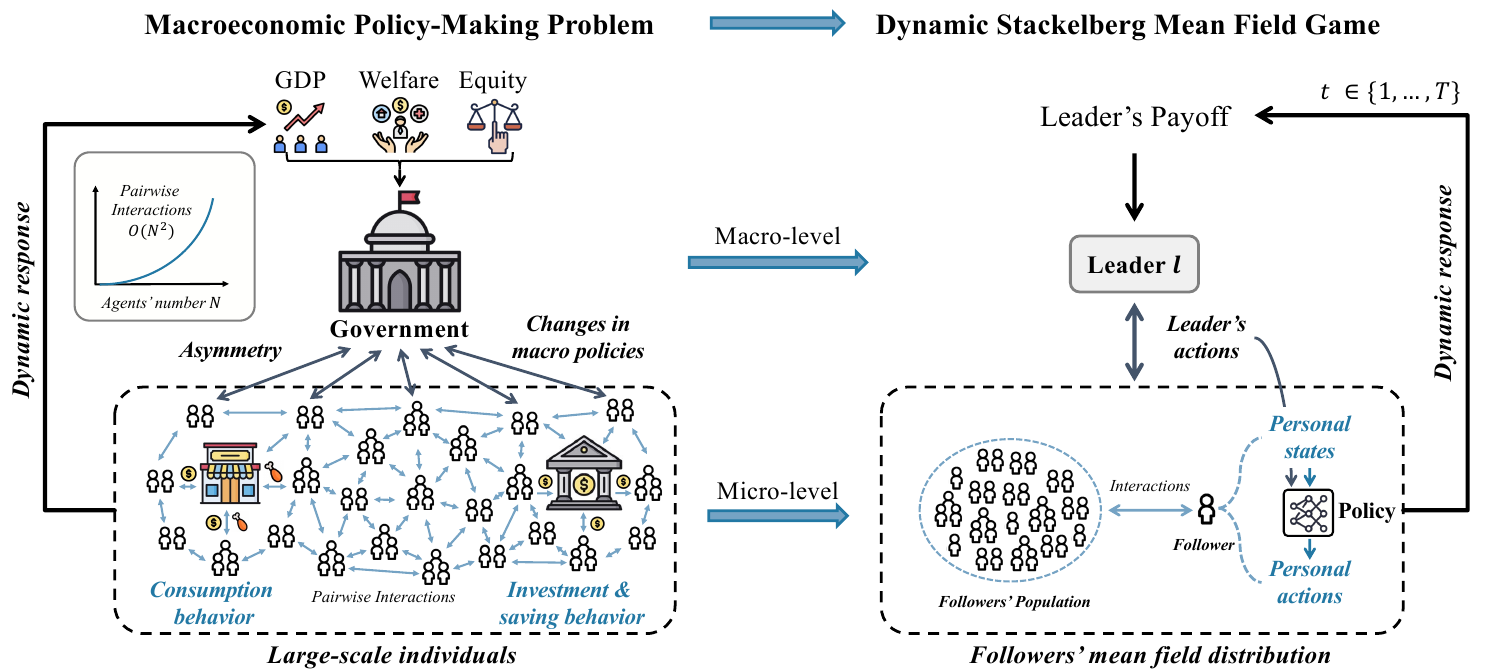}
    \vskip -0.1 in 
    \caption{
Macroeconomic policy-making involves both intensive individual–individual and individual–government interactions, whose pairwise complexity grows as \(O(N^2)\) with the agent number $N$—rendering direct simulation intractable. We address this by modeling the process as a \textbf{Dynamic Stackelberg Mean Field Game (DSMFG)}, which approximates these complex interactions via agent–population and leader–population couplings. This DSMFG retains key personal behaviors while enabling scalability and capturing three defining features of real-world policymaking: \textit{dynamics}, \textit{asymmetry}, and \textit{large-scale}.}
    \label{fig:SMFGl}
    \vskip -0.1 in 
\end{figure*}
\begin{itemize}[leftmargin=1.em,itemsep=0pt,topsep=0pt]
  \item A \textbf{scalable DSMFG framework} that integrates three core features—dynamic feedback, asymmetry, and large scale—into a unified model for macroeconomic policymaking. (Section~\ref{modeling})
  \item A \textbf{SMFRL algorithm} that efficiently learns macro policies under DSMFG while preserving personalized decision-making at the individual level. (Section~\ref{solution})
  \item \textbf{Extensive empirical validation} showing that SMFG scales to 1,000 agents and outperforms both economic and AI-based baselines. (Section~\ref{experiment})
\end{itemize}


\section{Related Work}

\subsection{Macroeconomic Models}
Classical macroeconomic frameworks—such as IS–LM~\cite{hicks1980lm,gali1992well}, AD–AS~\cite{dutt2006aggregate,lee1997growth}, and the Solow growth model~\cite{brock2010green}—have elucidated short- and long-run policy effects. New Keynesian DSGE models~\cite{blanchard2007real,smets2007shocks} then introduced micro individuals and stochastic shocks, enabling rigorous analysis under rational expectations. The Saez tax model~\cite{saez2001using} further offered a practical, elasticity-based tool for setting optimal tax rates. However, these approaches commonly rely on linearization, representative-agent assumptions, or fixed price-stickiness parameters, which prevent them from capturing individual responses, nonlinear feedback loops, and aggregate dynamics in large populations. Empirical and econometric methods~\cite{ramesh2010public,davidson2004econometric} quantify policy impacts from historical data but struggle with sparsity, identification, and out-of-sample validity.  
These gaps highlight the necessity of a framework that models dynamic interactions between the government and large-scale individuals.  

\subsection{RL for Economic Policy}
Recent advances have applied reinforcement learning (RL) into economic modeling. In macroeconomic settings, frameworks like AI Economist~\cite{zhengAIEconomistTaxation2022} use curriculum learning to optimize tax schedules, and others apply RL to crisis management~\cite{trott2021buildingfoundationdatadriveninterpretable}, monetary policy design~\cite{hinterlangOptimalMonetaryPolicy2021,chenDeepReinforcementLearning2023}, international trade dynamics~\cite{sch2021intelligence}, and market pricing with externalities~\cite{danassisAIdrivenPricesExternalities2023}. These studies typically treat the government as an independent decision-maker, overlooking how heterogeneous households adapt over time.  
At the micro level, RL has been used to study optimal savings and consumption~\cite{shiCanAIAgent2021, ruiLearningZeroHow2022, atashbarAIMacroeconomicModeling2023}, solve heterogeneous general-equilibrium problems~\cite{kurikshaEconomyNeuralNetworks2021,hill2021solvingheterogeneousgeneralequilibrium}, and model agent behaviors in barter~\cite{johansonEmergentBarteringBehaviour2022} and asset allocation~\cite{ozhamaratliDeepReinforcementLearning2022a}. While these works showcase RL’s promise, they rely on simplified environments or small agent populations, limiting their applicability to dynamic, large-scale macroeconomic policy design.  

\subsection{Stackelberg Mean Field Games}
Stackelberg Mean Field Games (SMFGs) integrate Stackelberg leader–follower dynamics with mean-field approximations to model interactions in large populations. Early \emph{model-based} SMFG methods solve forward–backward stochastic differential equations to compute follower equilibria before optimizing the leader’s policy~\cite{fu2020mean,dayanikli2023machine,bergault2023mean}. Linear–quadratic formulations~\cite{bensoussan2017linear,moon2018linear,huang2020mean} and minimax rewritings~\cite{guo2022optimization} enhance analytical tractability but impose restrictive assumptions on dynamics and transitions, limiting applicability to complex economic settings.
\emph{Model-free} SMFG approaches dispense with explicit transition models by learning directly from interaction data. For example, \citet{pawlick2017mean} handle single-step SMFGs, \citet{campbell2021deep} apply deep BSDE solvers for equilibrium computation, and \citet{miao2024mean} explore defensive follower strategies under fixed attacker trajectories. More recently, \citet{li2024transition} estimate empirical transition kernels and solve the resulting Fokker–Planck equations. However, these methods remain confined to simplified benchmarks—single-period decisions, low-dimensional state spaces, or predefined follower classes—and do not capture the multi-period feedback loops and scale required for realistic macroeconomic policy design.
Thus, there is a clear need for a stable, model-free algorithm that can solve SMFGs under complex, dynamic economic interactions without requiring knowledge of true transition dynamics.


\section{Dynamic Stackelberg Mean Field Game Framework}\label{modeling}

In this section, we first identify the core features of the macroeconomic policy-making problem, and then propose a dynamic Stackelberg mean-field game framework to model them effectively.

\subsection{Macroeconomic Policy‐Making Problem}

Macroeconomic policy comprises government actions—such as monetary and fiscal interventions—that stabilize growth, reduce unemployment, and control inflation~\cite{barro1990macroeconomic}. These interventions shape individual decisions (e.g.\ labor supply, consumption, investment), which in turn generate aggregate outcomes that inform subsequent policy adjustments~\cite{miranda2021transmission}.
In the left panel of Figure~\ref{fig:SMFGl}, for instance, a change in the central bank’s interest-rate rule shifts households’ portfolio allocations, while each household’s choice also depends on the behavior of others. Such large-scale interactions induce complex feedback loops that static or small-scale models fail to capture.

We identify three salient, interdependent features of the macroeconomic policy-making problem:

\begin{enumerate}[leftmargin=1.5em,itemsep=0.1pt,topsep=0pt]
  \item \textbf{Dynamic feedback.}  
   A policy change triggers micro-level behavioral adjustments; the aggregate of these adjustments produces new macro indicators, which then feed back into the next policy decision. Modeling this continuous loop is essential, yet beyond the scope of static modeling methods.
  \item \textbf{Asymmetry.}  
   The government (leader) first commits to a policy rule; individual agents (followers) observe this policy and then optimize their private objectives. This sequential leader-follower structure underlies the inherent asymmetry dynamics between policymaker and population.
  \item \textbf{Large scale. } 
   Effective macro policy influences large scale micro-agents, thereby rendering the dynamic, asymmetric interactions described above more complex.
\end{enumerate}
\subsection{Dynamic Stackelberg Mean Field Game}
To design effective macroeconomic policies, we must first model three core features—dynamic feedback, asymmetry, and large scale—introduced above.

To capture \textbf{dynamic feedback}, we model macroeconomic policymaking as a dynamic game in which the government and individuals iteratively update their strategies based on evolving economic conditions. These strategies influence both immediate outcomes and the future decisions of other players~\cite{haurie2012games, van2010survey}.

To capture \textbf{asymmetry}, we adopt a Stackelberg leader–follower framework~\cite{simaan1973stackelberg}, modeling the government as the leader that sets policies first, followed by individuals’ responses.

To capture the \textbf{large-scale} nature of macroeconomic systems, an agent-based model with numerous micro-agents could be considered. However, scaling agent-based models to large populations is computationally challenging. With \(N\) agents, the model requires \(O(N^2)\) pairwise agent–agent interactions and \(O(N)\) government–agent interactions, rendering it computationally intractable for large-scale systems. To address this challenge, we employ a mean-field approximation~\cite{lasry2007mean}, where interactions between individual agents are replaced by those between a representative agent and the aggregate population, and government–agent interactions are modeled as interactions with the population’s mean field. This approach reduces the computational complexity from \(O(N^2)\) to \(O(N)\), significantly streamlining policy optimization.

In summary, our \textbf{Dynamic Stackelberg Mean-Field Game (DSMFG) framework} effectively integrates the three core features—\textit{dynamic feedback, asymmetry, and large scale}—into a cohesive model for macroeconomic policymaking. The precise mathematical formulation of DSMFG is as follows:

\paragraph{Framework Overview  }
In the DSMFG framework, we consider one leader and \(N\) follower agents. At each time step \(t \in \{0, \ldots, T\}\), the leader selects an action \(a_t^l \in \mathcal{A}^l\) based on its state \(s_t^l \in \mathcal{S}^l\) and a policy \(\pi^l: \mathcal{S}^l \to \mathcal{A}^l\). Subsequently, the followers determine their actions based on the leader's action \(a_t^l\) and their private states \(s_t^f \in \mathcal{S}^f\). A representative follower’s action \(a_t^f \in \mathcal{A}^f\) is derived from a shared policy \(\pi^f: \mathcal{S}^f \times \mathcal{A}^l \to \mathcal{A}^f\). The sequences \(\{\pi_t^l\}_{t=0}^T\) and \(\{\pi_t^f\}_{t=0}^T\) are denoted as \(\pi^l\) and \(\pi^f\), respectively. The mean field \(L_t(s_t^f, a_t^f; \pi^f, a_t^l)\), abbreviated as \(L_t(s_t^f, a_t^f)\), represents the population state-action distribution of followers, defined as:
\[
    L_t(s_t^f, a_t^f; \pi^f, a_t^l) \in \mathcal{P}(\mathcal{S}^f \times \mathcal{A}^f), \text{ where } a_t^f = \pi^f(s_t^f, a_t^l).
\]

\paragraph{Followers}\label{fmfg}
At each time step \(t \in \{0, \ldots, T-1\}\), given the joint state \(\mathbf{s}_t = \{s_t^l, \mathbf{s}_t^f\}\), a representative follower receives a reward \(r^f(\mathbf{s}_t, a_t^l, a_t^f, L_t)\) and transitions to the next state \(s_{t+1}^f \sim P(\cdot \mid \mathbf{s}_t, a_t^l, a_t^f, L_t)\).
The follower’s objective is to optimize their policy \(\pi^f\) to maximize cumulative rewards over the time horizon:
\[
    J^f(\pi^l, \pi^f, L) = \mathbb{E}_{s_0^f \sim \mu_0^f, \mathbf{s}_{t+1} \sim P} \left[\sum_{t=0}^T r^f(\mathbf{s}_t, a_t^l, a_t^f, L_t)\right],
\]
where the leader’s action \(a_t^l = \pi^l(s_t^l)\), the follower’s action \(a_t^f = \pi^f(s_t^f, a_t^l)\), and the mean field \(L_t = L_t(s_t^f, a_t^f)\).

\begin{definition}
    [Followers' Best Response for Leader's Policy]  
    Given a leader's policy \(\pi^l \in \Pi^l\) and followers' state-action distributions \(L = \{L_t\}_{t=0}^T\), the followers' best response policy \({\pi^f}^*(\pi^l, L)\) is defined as:
    \[
        {\pi^f}^*(\pi^l, L) \in \arg\max_{\pi'} J^f(\pi^l, \pi', L).
    \]\label{br_follower}
\end{definition}
\paragraph{Leader} 
At each time step \(t \in \{0, \ldots, T-1\}\), the leader receives a reward \(r^l(\mathbf{s}_t, a_t^l, L_t)\) based on its state \(s_t^l\), action \(a_t^l\), and the population mean field \(L_t\), and transitions to the next state $s_{t+1}^l \sim P(\cdot \mid \mathbf{s}_t, a_t^l, L_t)$.
The leader aims to optimize its policy \(\pi^l\) to maximize the expected cumulative reward:
\begin{equation}
    \begin{aligned}
        J^l(\pi^l, \pi^f, L)=\mathbb{E}_{s_0^l \sim \mu_0^l, \mathbf{s}_{t+1} \sim P} \left[\sum_{t=1}^T r^l(\mathbf{s}_t, a_t^l, L_t) \right],
    \end{aligned} \label{leader_utility}
\end{equation}
where $a^l_t=\pi^l(s^l_t)$, $a^f_t=\pi^f(s^f_t,a^l_t)$ and $L_t=L_t(s^f_t,a^f_t)$.

\begin{definition}
    [Leader's Optimal Policy in Dynamic Stackelberg Mean Field Games] Considering the followers' best response $(\pi^f, L)$ to the leader's policy, which satisfies Definition~\ref{br_follower}, learning the leader's optimal policy ${\pi^l}^*$in dynamic Stackelberg mean field games is equivalent to solving the following fixed-point problem given initial condition $(\mu^l_0, \mu^f_0)$:
\begin{equation*}
\begin{aligned}
    {\pi^l}^* &\in \mathop{\arg\max}_{{\pi^l}'} J^l({\pi^l}', \pi^f, L)\\
    &\text{s.t. } \pi^f \in \arg\max_{\pi'}J^f({\pi^l}^*, \pi', L)
\end{aligned}
\end{equation*}\label{br_leader}
where the followers' state-action distribution $L_t$ satisfies the following McKean-Vlasov equation:
\begin{equation*}
    \begin{aligned}
        L_{t+1}(s_{t+1}^f, a_{t+1}^f)=  \sum_{s_t^l, a_t^l, s_t^f, a_t^f}L_{t}(s_{t}^f, a_{t}^f)\pi_{t}^l(a_t^l \mid s_t^l)\mu_{t}^l(s_t^l)  \\
        P(s_{t+1}^f \mid s_{t}^f, a_{t}^f, a_{t}^l, L_t)\pi_{t+1}^f(a_{t+1}^f \mid s_{t+1}^f),
    \end{aligned}
\end{equation*}
\begin{equation*}
    \mu_{t+1}^l(s_{t+1}^l)=\sum_{s_t^l, a_t^l}\mu_t^l(s_t^l)\pi_t^l(a_t^l \mid s_t^l)P(s_{t+1}^l \mid s_t^l, a_t^l, L_t).
\end{equation*}
\end{definition}
In conclusion, we model the problem of macroeconomic policy-making as a DSMFG, capturing three core features. Based on this model, we optimize the government's policy by taking into account the followers’ best responses over discrete timesteps (Definition~\ref{br_leader}).

\section{Stackelberg Mean Field Reinforcement Learning}\label{solution}

\begin{figure}[t]
    \centering
    \includegraphics[width=0.85\linewidth]{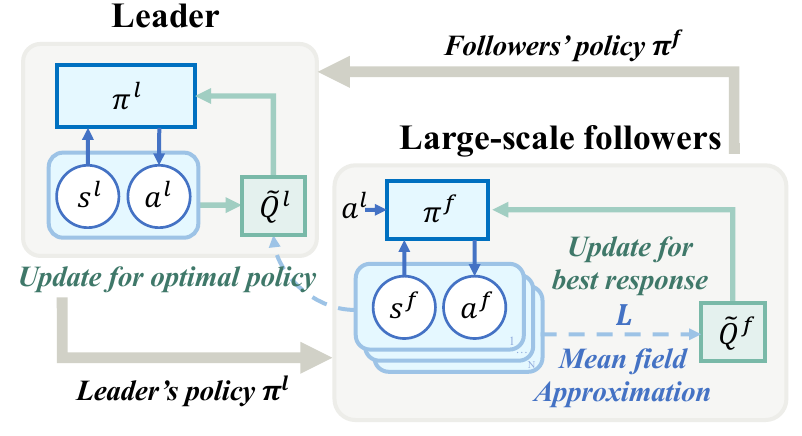}
    \caption{The architecture of SMFRL algorithm.}
    \label{smfrl}
\end{figure}

Within the DSMFG framework, we propose the \textbf{Stackelberg Mean Field Reinforcement Learning (SMFRL)} algorithm (Figure~\ref{smfrl}), which adopts a centralized training with decentralized execution (CTDE) paradigm~\cite{lowe2017multi}.
In standard CTDE settings, the leader agent is equipped with a policy \(\pi^l(s^l)\) and a centralized critic \(Q^l(s^l, a^l, \mathbf{s}^f, \mathbf{a}^f)\), while each follower is equipped with a policy \(\pi^f(s^f, a^l)\) and a corresponding Q-function \(Q^f(s^l, a^l, \mathbf{s}^f, \mathbf{a}^f)\). However, the joint state \(\mathbf{s}^f\) and action \(\mathbf{a}^f\) of all followers scale linearly with the population size, rendering direct learning of \(Q^l(s^l, a^l, \mathbf{s}^f, \mathbf{a}^f)\) computationally infeasible in large-scale environments. This directly reflects the computational complexity challenge induced by large-scale agent interactions, as discussed in Section~\ref{modeling}.

\subsection{Stackelberg Mean-Field Q and Policy} \label{mfa}
\paragraph{Stackelberg Mean-Field Q}
Within the DSMFG framework, we abstract the interactions between the leader and all followers into an interaction between the leader and the population mean field, thereby the leader's centralized Q-function can be reformulated as:
\[
Q^l(s^l, a^l, \mathbf{s}^f, \mathbf{a}^f) \approx \tilde{Q}^l(s^l, a^l, L).
\]

Similarly, based on the mean field approximation, we simplify the interaction of a follower with the leader and other followers into an interaction with the leader and the population mean field. Accordingly, the original Q-function \(Q^f(s^l, a^l, \mathbf{s}^f, \mathbf{a}^f)\) can be approximated as:
\begin{equation*}
    \begin{aligned}
        Q^f(s^l, a^l, \mathbf{s}^f, \mathbf{a}^f) &= Q^f(s^l, a^l, s^{fi}, a^{fi}, \mathbf{s}^{f,-i}, \mathbf{a}^{f,-i}) \\
        &\approx \tilde{Q}^f(s^l, a^l, s^{fi}, a^{fi}, L),
    \end{aligned}
\end{equation*}
where \((s^{fi}, a^{fi})\) denotes the individual state-action pair of the \(i\)-th follower, and \((\mathbf{s}^{f,-i}, \mathbf{a}^{f,-i})\) represent the rest of the population. Given that each individual has a negligible impact on the collective, the population mean field \(L\) can be used to approximate the rest of the population, following the theory of mean field games~\cite{lasry2007mean}.

In experiments, the mean field \(L\) is constructed from the empirical distribution over the followers' state-action pairs. At timestep \(t\),
\[
L_t = P(s^f, a^f), \quad (s^f, a^f) \sim \{(s^{f_i}_t, a^{f_i}_t)\}_{i=1}^N,  a^{f_i}_t \sim \pi^f(\cdot \mid s^{f_i}_t, a^l_t),
\]
where \(N\) denotes the number of followers.

In prior works on mean-field methods, a widely used simplification is to approximate \(L_t\) using population averages:
\begin{equation}
    L_t \approx (\bar{s}^f, \bar{a}^f), \quad \bar{s}^f = \frac{1}{N} \sum_{i=1}^N s^{f_i}, \quad \bar{a}^f = \frac{1}{N} \sum_{i=1}^N a^{f_i}.
\label{L}
\end{equation}
Alternative representations include neighborhood-based action averages~\cite{yang2018mean}, empirical distributions~\cite{carmona2018probabilistic}, and graph-based weighted mean fields~\cite{hao2023gat}. In our experiments, we find that using the average-based mean field achieves strong empirical performance while maintaining tractable computational cost.

In reinforcement learning, the local Q-functions \(\tilde{Q}^l\) and \(\tilde{Q}^f\) are defined as the expected cumulative rewards under discount factor \(\gamma \in [0, 1]\), starting from given states and actions:

\begin{equation}
    \begin{aligned}
    \tilde{Q}^l(s^l, a^l, s^f, a^f) = \mathbb{E} \left[ \sum_{t=0}^T \gamma^t r_t^l  \right],  \\
\tilde{Q}^f(s^l,a^l, s^{fi}, a^{fi}, s^{f},a^{f} ) = \mathbb{E} \left[ \sum_{t=0}^T \gamma^t r^{fi}_t\right].
    \end{aligned}
\end{equation}
Here, \(r_t^l\) and \(r_t^{fi}\) are the rewards for the leader and $i$-th follower, respectively. The discount factor \(\gamma\) determines the planning horizon: \(\gamma = 0\) models myopic agents focusing on immediate rewards, while \(\gamma > 0\) models non-myopic agents with long-term effects, increasing computational complexity.

\paragraph{Followers' policy}
Under the standard CTDE paradigm, the leader's policy takes the individual state as input, while the follower's policy conditions on both the individual's state and the leader's action. These input dimensions are tractable. However, training a separate policy for each follower is computationally infeasible.

To address this, we adopt a shared policy \(\pi^f(s^f, a^l)\) for all followers, as required by the mean-field setting. While this introduces inherent homogeneity—a known limitation of mean-field methods—\textit{we preserve individual heterogeneity by encoding personalized information in the state input}.

In reinforcement learning, a policy maps states to actions. Through training, the model learns how state features influence decisions. For instance, individual states include attributes such as age, education, and wealth, while actions cover economic choices like investment. This enables the shared policy to generate personalized behaviors, e.g., individuals of different ages and wealth levels exhibit distinct investment patterns.

\subsection{Leader-follower Update}\label{lf_update}
To enhance the convergence and stability, we propose the leader-follower update (shown in Figure~\ref{smfrl}): first, by fixing the leader's policy, we train the followers' shared policy and Q-networks towards the best response; subsequently, based on the followers' policies, we optimize the leader's policy and Q net, alternating these steps until convergence. We measure the distance between the agents' policies and their best responses by using exploitability.

Based on mean field approximation, we will train these networks \(\pi_{\theta^l}\) and \(\tilde{Q}_{\phi^l}\) for the leader agent, shared \(\pi_{\theta^f}\) and \(\tilde{Q}_{\phi^f}\) for the follower agents, with parameters \(\theta^l\), \(\phi^l\), \(\theta^f\), and \(\phi^f\). To ensure training stability, we introduce target networks with parameters \(\theta^l_{-}\), \(\phi^l_{-}\), \(\theta^f_{-}\), and \(\phi^f_{-}\). At any step \(t \in \{0,\ldots,T\}\), the tuple \((s^l_t, a^l_t, \mathbf{s}^f_t, \mathbf{a}^f_t, s^l_{t+1}, \mathbf{s}^f_{t+1}, r^l_t,\mathbf{r}^f_t)\) is stored in the replay buffer \(\mathcal{D}\) for training. 

\paragraph{Followers' Update for Best Response} 
Given leader's policy $\pi_{\theta^l}$, the followers' policy network $\pi_{\theta^f}$ is updated using a deterministic policy gradient \cite{silver2014deterministic}. The followers' policy gradient is estimated:
\begin{equation*}
\begin{aligned}
   \nabla_{\theta^f} J \approx \mathbb{E}_{\mathbf{s}_t, a^l_t, L_t\sim \mathcal{D} }
    \Big[ \nabla_{\theta^f} \pi_{\theta^f}(s^{f}_t, a^l_t) \nabla_{a^{f}_{-}} \tilde{Q}_{\phi^f}(s^l_t, a^l_t, s^{f}_t,a^{f}_{-},L_t)\Big]
\end{aligned}
\end{equation*}
where $a^{f}_{-}=\pi_{\theta^f}(s^{f}_t, a^l_t)$, $L_t$ is computed by state-action pairs sampled from replay buffer $\mathcal{D}$ by Eq.~(\ref{L}). The action-value function $\tilde{Q}_{\phi^f}$ is updated by minimizing the mean squared error loss:
\begin{equation*}
\small
\begin{aligned}
\mathcal{L}(\phi^f) = \mathbb{E}_{\substack{\mathbf{s}_t,a^l_t, L_t,\mathbf{s}_{t+1} \sim \mathcal{D} }} \left[\left(y^{f}_t - \tilde{Q}_{\phi^f}(s^l_t, a^l_t, s^{f}_t,a^{f}_{-},L_t) \right)^2\right]
\end{aligned}
\end{equation*}
\begin{equation*}
\small
\begin{aligned}
    y^{f}_t = r^{f}_t +  \gamma\tilde{Q}_{\phi^f_-}(s^l_{t+1}, a^l_{t+1}, s^{f}_{t+1},a^{f}_{t+1},L_{t+1})
\end{aligned}  
\end{equation*}
where $a^l_{t+1} = \pi_{\theta^l}(s^l_{t+1})$, $a^{f}_{t+1}=\pi_{\theta^f_{-}}(s^{f}_{t+1}, a^l_{t+1})$.
The gradient of the loss function $\mathcal{L}(\phi^f)$ is derived as:
\begin{equation*}
\small
\begin{aligned}
    \nabla_{\phi^f} \mathcal{L}(\phi^f)=
    \mathbb{E}_{\substack{\mathbf{s}_t,a^l_t, L_t, \mathbf{s}_{t+1} \sim \mathcal{D} }}& \Big[\left(y^{f}_t - \tilde{Q}_{\phi^f}(s^l_t, a^l_t, s^{f}_t,a^{f}_{t},L_t) \right)  \\
    &\nabla_{\phi^f}\tilde{Q}_{\phi^f}(s^l_t, a^l_t, s^{f}_t,a^{f}_{t},L_t) \Big].
\end{aligned}
\end{equation*}

\paragraph{Leader's Update for Optimal Policy} 
Given followers' policy $\pi_{\theta^f}$, the leader's policy $\pi_{\theta^l}$ is optimized by DPG approach, and the leader's policy gradient is estimated:

\begin{equation*}
    \begin{aligned}
        \nabla_{\theta^l} J \approx \mathbb{E}_{\mathbf{s}_t, L_t\sim \mathcal{D}}  \left[\nabla_{\theta^l} \pi_{\theta^l}(s^l_t) \nabla_{a^l_{-}} \tilde{Q}_{\phi^l}(s_t^l,a^l_{-}, L_t) \right]|_{a^l_{-}=\pi_{\theta^l}(s^l_t)}.
    \end{aligned}
\end{equation*}
This network is periodically updated to minimize the loss:
\begin{equation*}
\begin{aligned}
\mathcal{L}(\phi^l) = \mathbb{E}_{\substack{\mathbf{s}_t,a^l_t, L_t,\mathbf{s}_{t+1} \sim \mathcal{D}}} \left[ \left(y^l_t - \tilde{Q}_{\phi^l}(s^l_t, a^l_t, L_t) \right)^2 \right]
\end{aligned}
\end{equation*}
The target value $y^l_t$ is given by:
\begin{equation*}
\small
\begin{aligned}
    y^l_t = r_t^l + \gamma  \tilde{Q}_{\phi^l_{-}}(s^l_{t+1},a^l_{t+1},L_{t+1})|_{a^l_{t+1} = \pi_{\theta^l_{-}}(s^l_{t+1})}
\end{aligned}
\end{equation*}
where $\gamma$ is the discount factor. Differentiating the loss function $\mathcal{L}(\phi^l)$ yields the gradient utilized for training: 
\begin{equation*}
\begin{aligned}
    \nabla_{\phi^l} \mathcal{L}(\phi^l)=
    \mathbb{E} \left[ \left(y^l_t - \tilde{Q}_{\phi^l}(s^l_t, a^l_t, L_t) \right)
    \nabla_{\phi^l}\tilde{Q}_{\phi^l}(s^l_t, a^l_t,L_t) \right].\label{leader_critic}
\end{aligned}
\end{equation*}
where the expectation $\mathbb{E}$ is taken over \((\mathbf{s}_t, a^l_t, L_t, \mathbf{s}_{t+1}) \sim \mathcal{D}\).
The pseudocode for the SMFRL algorithm~\ref{alg:smfrl} in Appendix~\ref{appendix_smfrl}.

\section{Experiment}\label{experiment}

\begin{table*}[h]
\centering
\caption{Performance of multiple policies on key macroeconomic indicators for $N=100$ and $N=1000$ households. The best values are highlighted in \textbf{bold}, and the second-best values are \underline{underlined}.}
\vskip -0.1in
\label{tab:policy_comparison}
\resizebox{\textwidth}{!}{%
\begin{tabular}{ll l c c c c c cc c}
\toprule
\textbf{Category}    & \textbf{Subcategory}  & \textbf{Policies}        & \multicolumn{2}{c}{\textbf{Per Capita GDP $\uparrow$}} & \multicolumn{2}{c}{\textbf{Social Welfare $\uparrow$}} & \multicolumn{2}{c}{\textbf{Wealth Gini $\downarrow$}} & \multicolumn{2}{c}{\textbf{Years $\uparrow$}} \\ 
\cmidrule(lr){4-5} \cmidrule(lr){6-7} \cmidrule(lr){8-9} \cmidrule(lr){10-11}
 &  &         & $100$ & $1000$ & $100$ & $1000$ & $100$ & $1000$ & $100$ & $1000$ \\ 
\midrule
\multirow{1}{*}{Static} 
& Non-intervention        & \textbf{Free Market} & 1.37e+05 & 1.41e+05 & 32.97  & 334.79 & 0.92  & 0.93 & 1.10  & 1.00\\ 
\multirow{1}{*}{Policies} 
& Real-data              & \textbf{US Federal Tax} & 4.88e+11 & 1.41e+05 & 94.19  & 351.17 & \underline{0.40}  & 0.93 & 289.55 & 1.00 \\ 
 \midrule
 \multirow{3}{*}{Dynamic} 
& Rule-based    & \textbf{Saez Tax} & \underline{2.34e+12} & \underline{6.35e+11 }& 73.82  & 498.88 & \textbf{0.38}  & \underline{0.73} & \textbf{300.00} & \underline{100.58} \\  
 \multirow{3}{*}{Policies} 
& Independent-based        & \textbf{AI Economist} & 1.26e+05 & {N/A} & 72.81  & {N/A} & 0.91  & {N/A} & 1.00  & {N/A} \\ 
& \multirow{4}{*}{Game-based } 
  & \textbf{DSMFG (ours)} & \textbf{9.59e+12} & \textbf{1.10e+13} & \textbf{96.87}  & \textbf{968.94} & {0.51}  & \textbf{0.53} & \textbf{300.00} & \textbf{300.00} \\ 
  &    & \textbf{DSMFG w/o S} & 8.66e+07 & 1.31e+05 & 82.02  & 834.93 & 0.83  & 0.92 &\underline{ 75.75}  & 1.50 \\ 
&  & \textbf{DSMFG w/o MF} & 1.23e+05 & 1.48e+05 & 48.17  & 499.01 & 0.93  & 0.93 & 1.00  & 1.02 \\ 
& &  \textbf{DSMFG w/o MF \& S}     & 1.21e+05 & 1.33e+05 & \underline{83.09}  & \underline{874.53} & 0.92  & 0.92 & 1.00  & 1.00 \\ 

\bottomrule
\end{tabular}%
}
\end{table*}
To validate the effectiveness of our DSMFG framework and SMFRL algorithm for macroeconomic policymaking, our experiments are designed to answer the following key questions:
\begin{enumerate}[leftmargin=*, itemsep=0pt, topsep=2pt]
    \item \textbf{Effectiveness of dynamic modeling.} Does the DSMFG framework outperform static macroeconomic policies in optimizing critical economic indicators? (\S~\ref{experiment_dynamic_game})
    \item \textbf{Necessity of Stackelberg structure and mean field approximation.} Are the Stackelberg structure and mean field approximation essential for the scalability and performance of DSMFG? (\S~\ref{SMFG_importance}) 
    \item \textbf{Impact of mean field homogeneity.} How does the homogeneity assumption inherent in mean field approximations affect decision personalization and policy robustness? (\S~\ref{heter})
\end{enumerate}
In the appendix, we include details on computational resources and efficiency (\ref{resource}), full training curves and result tables (\ref{smfg_nece},\ref{training_tax}), and discussions on the efficiency-equity of different policies (\ref{tradeoff}).

\subsection{Experimental Setting}
\paragraph{Environment}
We conduct experiments in TaxAI~\cite{mi2023taxai}, a simulation platform for optimal tax policy. TaxAI enables dynamic interactions between governments and large-scale households using \textbf{real-world datasets}. Details are in Appendix~\ref{intro_taxai}.

\paragraph{Evaluation Metrics}
 \textit{Per Capita GDP} reflects the level of economic development, while \textit{Income Gini} and \textit{Wealth Gini} measure inequality in household income and wealth, respectively—a lower Gini index indicates greater social equality. The \textit{Years} metric represents the sustainable duration of an economy, with a maximum cap of 300 years. \textit{Average Wealth}, \textit{Income}, and \textit{Consumption} are crucial assessment metrics related to financial crises. 
\paragraph{Baselines}
We compare our method against static, dynamic, and game-based policies to validate the necessity of the proposed DSMFG framework (see Table~\ref{tab:policy_comparison}). The parameters of the baselines are provided in Appendix~\ref{parameters}. 
\begin{itemize}[itemsep=0pt]
    \item \textbf{Static Policies:}
    \begin{itemize}[itemsep=0pt]
        \item \textbf{Free Market}~\cite{backhouse2005rise}: No government intervention.
        \item \textbf{U.S. Federal Tax}: The actual progressive personal income tax policy implemented by the U.S. federal government in 2022. This serves as a strong static policy baseline.
    \end{itemize}

    \item \textbf{Dynamic Policies:}
    \begin{itemize}[itemsep=0pt]
        \item \textbf{Saez Tax}~\cite{saez2001using}: A rule-based economic method widely recommended for tax reforms in real world (details in Appendix~\ref{saez}).
        \item \textbf{AI Economist}~\cite{zhengAIEconomistTaxation2022}: An independent-based policy employing independent Proximal Policy Optimization (PPO), which does not consider multi-agent interactions.
    \end{itemize}
    
    \item \textbf{Game-based Policies:}
    \begin{itemize}[itemsep=0pt]
        \item \textbf{DSMFG (Ours)}: Incorporates dynamic Stackelberg Mean Field Games.
        \item \textbf{DSMFG w/o S}: DSMFG without Stackelberg structure.
        \item \textbf{DSMFG w/o MF}: DSMFG without mean field approximation.
        \item \textbf{DSMFG w/o MF \& S}: DSMFG excluding both Stackelberg structure and mean field approximation.
    \end{itemize}
\end{itemize}

\subsection{Effectiveness of Dynamic Modeling}\label{experiment_dynamic_game}
\begin{figure*}[h]
    \centering
    \includegraphics[width=1\linewidth]{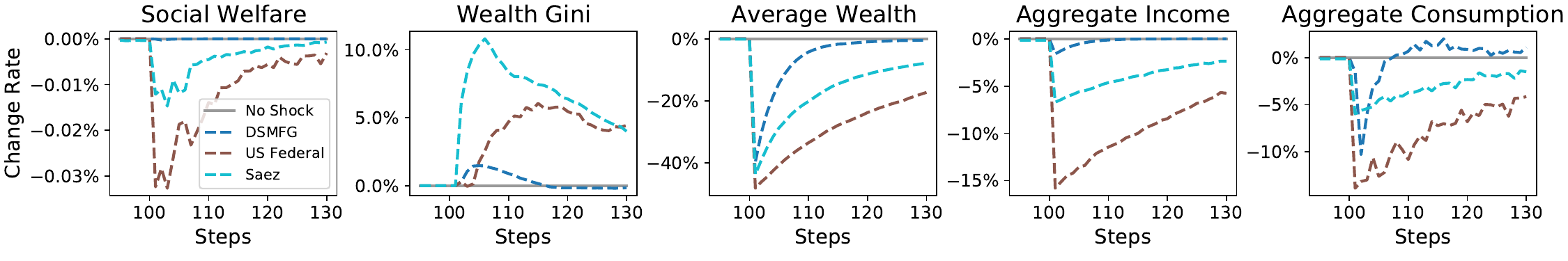}
    \vskip -0.12in
    \caption{Dynamic response curves of 3 macroeconomic policies to economic shocks at step 100. The DSMFG policy (dark blue line) exhibits the least fluctuation and the fastest recovery in key indicators, indicating superior dynamic response capabilities.}
    \label{crisis}
    \vskip -0.1in
\end{figure*}

Our DSMFG framework, a dynamic game-based policy, outperforms static, rule-based, and independent policies across key economic metrics (Table~\ref{tab:policy_comparison}) and responds more rapidly to economic shocks (Figure~\ref{crisis}).

\paragraph{Superior Performance of Dynamic Policies}
Table~\ref{tab:policy_comparison} ranks policy performance: DSMFG (dynamic game-based) $\succ$ Saez Tax (dynamic rule-based) $\succ$ U.S. Federal Tax (static) $\succ$ others. In the TaxAI environment, which models complex economic interactions, ineffective policies often trigger termination conditions (e.g., extreme inequality, Gini $> 0.9$). Free Market and AI Economist policies, lacking government leadership or multi-agent modeling, fail to achieve sustainable outcomes, as evidenced by the \textit{Years} metric. Among sustainable policies (U.S. Federal Tax, Saez Tax, DSMFG), dynamic policies outperform static ones: DSMFG achieves a \textit{Per Capita GDP} 19 times higher than U.S. Federal Tax ($9.59 \times 10^{12}$ vs. $4.88 \times 10^{11}$) and 4 times higher than Saez Tax ($9.59 \times 10^{12}$ vs. $2.34 \times 10^{12}$). At $N=1000$, U.S. Federal Tax, based on 2022 static data, fails to sustain economic development, while Saez Tax shows declining \textit{Per Capita GDP}, \textit{Social Welfare}, and \textit{Gini} metrics. In contrast, DSMFG maintains higher \textit{Per Capita GDP} with comparable \textit{welfare} and \textit{Gini}, demonstrating superior scalability and performance over static, rule-based, and AI-based policies.

\paragraph{Rapid Response to Economic Shocks}
To assess dynamic responsiveness, we simulate a financial crisis in the TaxAI environment, where all households lose 50\% of their wealth at step 100 (see Appendix~\ref{economic_shock} for details). As Free Market and AI Economist policies are limited to one-year simulations (Table~\ref{tab:policy_comparison}), we compare DSMFG against U.S. Federal Tax and Saez Tax. Figure~\ref{crisis} shows DSMFG (dark blue line) recovering fastest across all metrics: \textit{Social Welfare} remains stable, \textit{Average Income} and \textit{Consumption} recover within 5 steps, and \textit{Wealth Gini} and \textit{Average Wealth} stabilize within 15 steps. This unmatched resilience highlights DSMFG’s precise modeling of follower decisions and superior adaptability.

\begin{figure*}[h]
    \centering
    \includegraphics[width=0.95\linewidth]{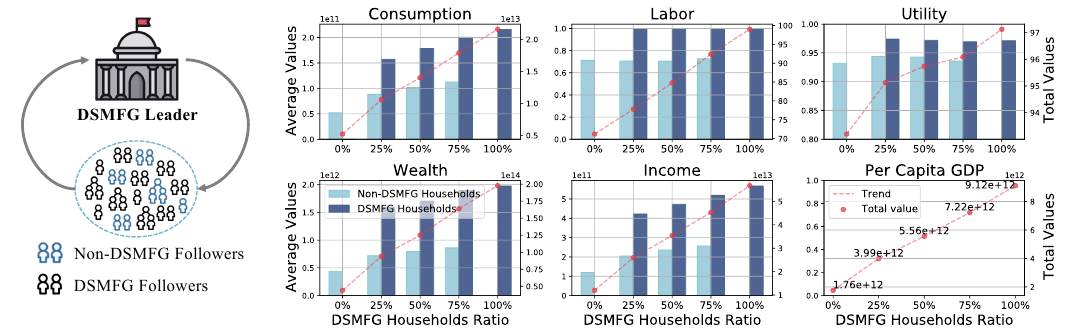}
    \vskip -0.15in
    \caption{The left subfigure illustrates a scenario with heterogeneous households adopting various policies. The right subfigure presents test results for households' decisions (consumption, labor), microeconomic indicators (utility, wealth, income), and macroeconomic indicators (GDP) when there are different proportions of DSMFG followers ($N=100$).}
    \label{fig:heter}
    \vskip -0.15in
\end{figure*}

\begin{figure*}[h]
    \centering
    \includegraphics[width=0.96\linewidth]{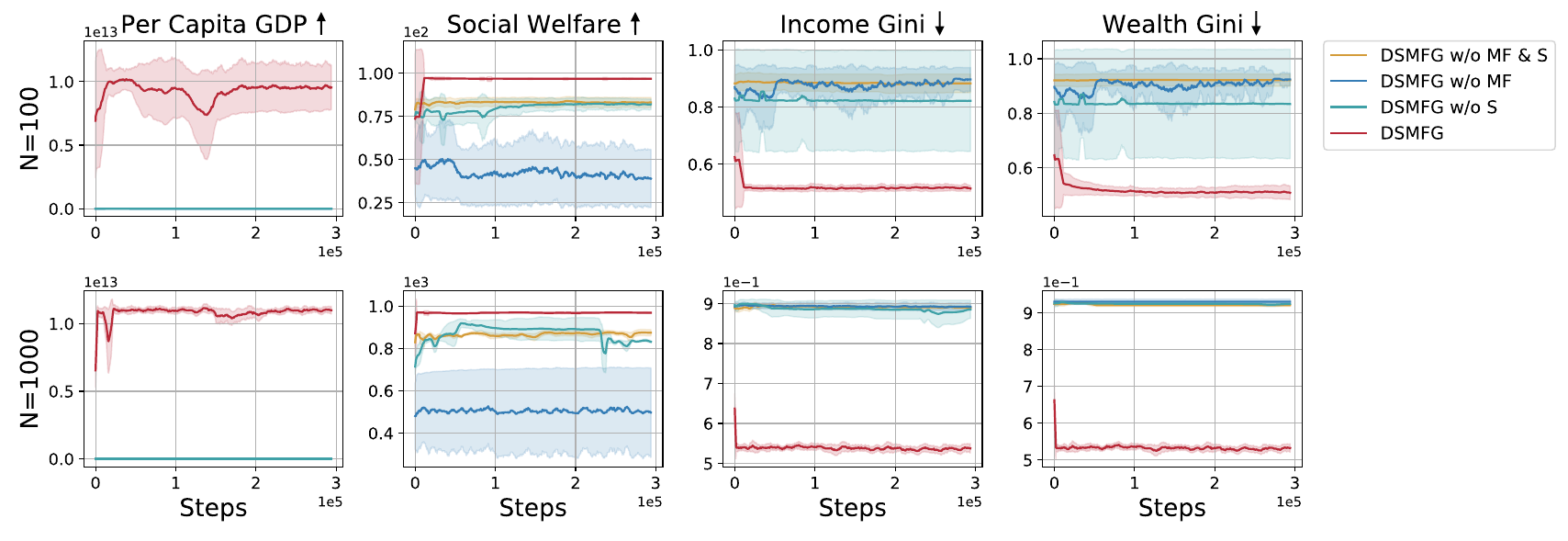}
    \vskip -0.15in
    \caption{The training curves for 6 algorithms on 4 macroeconomic indicators, comparing settings without behavior cloning as pre-train (N=100 \& N=1000) and with behavior cloning (N=100-BC \& N=1000-BC).}
    \label{fig:n_100_training}
    \vskip -0.1in
\end{figure*}
\subsection{Necessity of Stackelberg and Mean-Field Components}
\label{SMFG_importance}

To evaluate the importance of the Stackelberg and mean-field components in DSMFG, we compare DSMFG against its ablated variants (DSMFG w/o S, DSMFG w/o MF, DSMFG w/o MF \& S) using macroeconomic metrics (Table~\ref{tab:policy_comparison}), training dynamics (Figure~\ref{fig:n_100_training}), and game-theoretic indicators (Table~\ref{smfrl_table}). These indicators-leader payoff, exploitability, and social welfare-quantify the leader’s policy optimization, convergence to equilibrium, and follower policy quality, respectively (see Appendix~\ref{metrics} for details).

\paragraph{Economic performance and training stability}
Table~\ref{tab:policy_comparison} demonstrates DSMFG’s superior performance across all economic metrics. Removing either the Stackelberg or mean-field component causes substantial performance degradation. For instance, DSMFG w/o S yields a social welfare of $834.93$ for \(N=1000\), a $14\%$ drop from DSMFG, while DSMFG w/o MF reduces social welfare to $499.01$, nearly halving DSMFG’s value. Training curves in Figure~\ref{fig:n_100_training} further confirm these findings: DSMFG converges stably, while its variants exhibit slower convergence and higher variance (shaded regions, based on five seeds), particularly for DSMFG w/o S and DSMFG w/o MF. These results highlight the indispensable role of both components in achieving robust and optimal policy outcomes.

\begin{table}[ht]
\centering
\caption{Ablation studies of the DSMFG method based on game theory metrics for N=100 and N=1000. Optimal values are provided for reference.}
\vskip -0.1in
\begin{tabular}{lcccccc}
\toprule
\textbf{} & \multicolumn{2}{c}{\textbf{Leader's}} & \multicolumn{2}{c}{\textbf{Exploitability}} & \multicolumn{2}{c}{\textbf{Social}} \\
\textbf{} & \multicolumn{2}{c}{\textbf{Payoff}} & \multicolumn{2}{c}{} & \multicolumn{2}{c}{\textbf{Welfare}} \\
\cmidrule(lr){2-3} \cmidrule(lr){4-5} \cmidrule(lr){6-7}
\textbf{Methods} & 100 & 1000 & 100 & 1000 & 100 & 1000 \\
\midrule
{\textbf{Optimal Value}} & \textbackslash & \textbackslash & 0. & 0. & 100 & 1000 \\
{\textbf{DSMFG (ours)}} & \textbf{3294} & \textbf{3376} & \textbf{0.002} & \textbf{0.023} & \textbf{98} & \textbf{971} \\
{\textbf{DSMFG w/o S}} & -448 & -856 & 3.161 & 1.213 & 82 & 782 \\
{\textbf{DSMFG w/o MF}} & -535 & -441 & 0.782 & 0.725 & 54 & 499 \\
{\textbf{DSMFG w/o MF \& S}} & -774 & -716 & 0.652 & 1.023 & 83 & 859 \\
\bottomrule
\end{tabular}
\label{smfrl_table}
\vskip -0.1in
\end{table}

\paragraph{Game-Theoretic Analysis}
Table~\ref{smfrl_table} provides deeper insights into the contributions of each component. DSMFG achieves a leader payoff of $3294$ for \(N=100\) and an exploitability of $0.002$, closely approaching the optimal value of 0. \textbf{(i) Removing the Stackelberg module} (DSMFG w/o S) drastically reduces the leader payoff to $-448$ and increases exploitability to $3.161$, indicating the Stackelberg structure’s critical role in \textit{optimizing leader policies and ensuring equilibrium convergence}. Similarly, \textbf{(ii) removing the mean-field component} (DSMFG w/o MF) lowers social welfare to $54$ for \(N=100\), a $45\%$ reduction from DSMFG’s $98$, underscoring its necessity for \textit{effective follower policy optimization}. The combined ablation (DSMFG w/o MF \& S) reduces the framework to standard multi-agent reinforcement learning, performing comparably to DSMFG w/o MF but with increased exploitability ($1.023$ vs. $0.725$ for \(N=1000\)). This further emphasizes the Stackelberg module’s importance in large-scale settings, where it facilitates convergence to a stable equilibrium.

\subsection{Impact of Mean-Field Homogeneity}
\label{heter}
DSMFG leverages a mean-field approximation to ensure scalability in modeling and learning, but this introduces an inherent homogeneity assumption. In DSMFG, individual heterogeneity is preserved through state representations, allowing the shared policy to generate personalized actions. This section investigates the impact of shared follower policies by examining: (i) the effects of shared policies on training scalability and efficiency; and (ii) the robustness of DSMFG policies when interacting with heterogeneous follower behaviors.

\paragraph{Scalability and efficiency of shared policy}
The shared-policy design in DSMFG significantly improves both scalability and training efficiency compared to heterogeneous-policy variants, such as DSMFG w/o MF and DSMFG w/o MF \& S.  
\textbf{(i) Scalability:} As shown in Table~\ref{tab:policy_comparison}, heterogeneous-policy variants suffer from severe training instability, achieving sustainability for only $1$ year and producing lower GDP (\(1.48 \times 10^5\) vs. \(1.10 \times 10^{13}\) for DSMFG). Additionally, inequality worsens with a higher wealth Gini (0.93 vs. 0.53 at \(N=1000\)).  
\textbf{(ii) Efficiency:} Table~\ref{tab:experiment_results} (Appendix~\ref{resource}) shows that DSMFG reduces training time by 30\% to reach equivalent reward levels. These results highlight the advantages of shared-policy training in large-scale environments, enabling efficient convergence and better economic performance.

\paragraph{Robustness of DSMFG policies}
We test DSMFG policy robustness by introducing heterogeneous followers (named \textit{Non-DSMFG followers}) using behavior-cloned policies derived from real-world data (Appendix~\ref{bc_detail}). The leader and a subset of followers retain DSMFG-trained policies, while the rest adopt \emph{Non-DSMFG followers}. We vary the proportion of DSMFG followers (0\%, 25\%, 50\%, 75\%, 100\%) and track both micro-level (wealth, income, utility) and macro-level (GDP) indicators. Figure~\ref{fig:heter} reports average values (left Y-axis, bars) and total values (right Y-axis, points).

Results show three key findings:
(1) DSMFG followers consistently maintain high utility (96–97) across all proportions, indicating robustness against policy heterogeneity.  
(2) DSMFG followers outperform \emph{Non-DSMFG followers} in wealth, income, and utility—often by more than $\times 2$—demonstrating the superior effectiveness of DSMFG policies.  
(3) Per capita GDP increases monotonically with the proportion of DSMFG followers, suggesting that widespread adoption of DSMFG policies can yield substantial macroeconomic gains.

\section{Conclusion}
We introduce the Dynamic Stackelberg Mean Field Game (DSMFG) framework, which captures the dynamic, asymmetric, and large-scale nature of government–individual interactions in macroeconomic settings. To solve DSMFG, we develop the Stackelberg Mean Field Reinforcement Learning (SMFRL) algorithm, which combines Stackelberg game theory with mean-field approximation to enable scalable and efficient policy learning in large populations. This approach provides a principled and scalable solution to policy optimization problems that are otherwise computationally intractable.
Our results underscore the value of integrating game-theoretic modeling with data-driven learning for large-scale economic decision-making. Future directions include extending DSMFG to multi-policy settings, modeling richer behavioral heterogeneity, and calibrating with real-world economic data.






\bibliography{mybibfile}


\newpage
\clearpage

\appendix

\begin{algorithm*}[!ht]
   \caption{Stackelberg Mean Field Reinforcement Learning (SMFRL)}
   \label{alg:smfrl}
\begin{algorithmic}
\STATE Initialize $\tilde{Q}_{\phi^l}$, $\tilde{Q}_{\phi^l_-}$, $\tilde{Q}_{\phi^f}$, $\tilde{Q}_{\phi^f_-}$, $\pi_{\theta^l}$, $\pi_{\theta^l_-}$, $\pi_{\theta^f}$, $\pi_{\theta^f_-}$, replay buffer $\mathcal{D}$.
\FOR{epoch = 1 to M}
\STATE Receive initial state $\mathbf{s}_t=\{s^l_t, \mathbf{s}^f_t\}$.
\FOR{$t=1$ to max-epoch-length}
\STATE Leader action: $a^l_t=\pi_{\theta^l}(s^l_t) + \mathcal{N}_t$; followers' actions: $a^{f}_t=\pi_{\theta^f}(s^{f}_t, a^l_t) + \mathcal{N}_t$.
\STATE Execute $\mathbf{a}_t=\{a^l_t, \mathbf{a}^f_t\}$, observe rewards $\mathbf{r}_t$ and next state $\mathbf{s}_{t+1}$.
\STATE Store tuple $(s^l_t, a^l_t, \mathbf{s}^f_t, \mathbf{a}^f_t, s^l_{t+1}, \mathbf{s}^f_{t+1}, r^l_t,\mathbf{r}^f_t)$ in $\mathcal{D}$.
\STATE $\mathbf{s}_t \leftarrow \mathbf{s}_{t+1}$.
\FOR{j = 1 to update-cycles}
\STATE Sample minibatch from $\mathcal{D}$.
\STATE \textbf{Followers' Update:}
\STATE Compute follower targets: 
\[
y^{f}_t = r^{f}_t + \gamma \tilde{Q}_{\phi^f_-}(s^l_{t+1}, a^l_{t+1}, s^{f}_{t+1}, a^{f}_{t+1}, L_{t+1}) \mid_{ a^l_{t+1}=\pi_{\theta^l}(s^l_{t+1}), a^{f}_{t+1}=\pi_{\theta^f_-}(s^{f}_{t+1}, a^l_{t+1})}.
\]
\STATE Update follower critic $\tilde{Q}_{\phi^f}$ by minimizing:
\[\mathcal{L}(\phi^f)=\mathbb{E}\left[(y^{f}_t - \tilde{Q}_{\phi^f}(s^l_t, a^l_t, s^{f}_t,a^{f}_t,L_t))^2\right]\]
\STATE Update follower policy $\pi_{\theta^f}$ via:
\begin{equation*}
\begin{aligned}
   \nabla_{\theta^f} J \approx \mathbb{E}_{\mathbf{s}_t, a^l_t, L_t\sim \mathcal{D} }
    \Big[ \nabla_{\theta^f} \pi_{\theta^f}(s^{f}_t, a^l_t) \nabla_{a^{f}_{-}} \tilde{Q}_{\phi^f}(s^l_t, a^l_t, s^{f}_t,a^{f}_{-},L_t)\Big]
\end{aligned}
\end{equation*}
\STATE \textbf{Leader's Update:}
\STATE Compute leader target:
\[y^l_t=r_t^l+\gamma \tilde{Q}_{\phi^l_-}(s^l_{t+1},a^l_{t+1},L_{t+1}),\quad a^l_{t+1}=\pi_{\theta^l_-}(s^l_{t+1})\]
\STATE Update leader critic $\tilde{Q}_{\phi^l}$ by minimizing:
\[\mathcal{L}(\phi^l)=\mathbb{E}\left[(y^l_t - \tilde{Q}_{\phi^l}(s^l_t, a^l_t,L_t))^2\right]\]
\STATE Update leader policy $\pi_{\theta^l}$ via:
\begin{equation*}
    \begin{aligned}
        \nabla_{\theta^l} J \approx \mathbb{E}_{\mathbf{s}_t, L_t\sim \mathcal{D}}  \left[\nabla_{\theta^l} \pi_{\theta^l}(s^l_t) \nabla_{a^l_{-}} \tilde{Q}_{\phi^l}(s_t^l,a^l_{-}, L_t) \right]|_{a^l_{-}=\pi_{\theta^l}(s^l_t)}.
    \end{aligned}
\end{equation*}
\ENDFOR
\STATE Periodically update target networks:
\[
\phi_{-}^{l,f} \leftarrow \tau_\phi \phi^{l,f}+(1-\tau_\phi)\phi_{-}^{l,f},\quad
\theta_{-}^{l,f} \leftarrow \tau_\theta \theta^{l,f}+(1-\tau_\theta)\theta_{-}^{l,f}.
\]
\ENDFOR
\ENDFOR
\end{algorithmic}
\end{algorithm*}

\clearpage

\section{SMFRL algorithm pseudocode}\label{appendix_smfrl}
\section{Assumptions and Limitations}
\paragraph{Assumptions}\label{assumption}
This paper models the problem of macroeconomic policy-making as a Dynamic Stackelberg Mean Field Game, based on the following assumptions:
(1) Homogeneous followers: We assume that a large-scale group of households is homogeneous. They can use different characteristics as observations to influence decisions, but there are commonalities in human behavioral strategies.
(2) Rational Expectations: We assume that both macro and micro agents engage in rational decision-making, adjusting their future expectations based on observed information. However, in reality, the level of rationality varies among different households. Most households exhibit bounded rationality, and their expectations and preferences differ accordingly.
(3) Experimental environment: We validate our approach through experiments in the TaxAI environment, based on the assumption that results within this environment can provide insights applicable to real-world scenarios.
Addressing and potentially relaxing these assumptions will be a primary focus of our future research.

\paragraph{Limitations}\label{limitations}
The limitations of our DSMFG method will be thoroughly investigated in future work: 
(1) We plan to consider dynamic games involving multiple leaders and large-scale followers to explore policy coordination across various macroeconomic sectors.
(2) We will continue to develop theoretical proofs for the equilibrium solutions in Stackelberg mean field games. Currently, our approach is empirically demonstrated by showing that followers converge toward their best responses and that the leader achieves higher performance compared to other baselines. 
(3) We intend to examine dynamic games between a leader and a large, heterogeneous group of followers, including scenarios where followers dynamically alter their strategies, to determine the optimal leader policy. Addressing these limitations will provide further insights applicable to real-world scenarios.

\section{Saez tax}\label{saez}
The Saex tax policy is often considered a suggestion for specific tax reforms in the real world. The specific calculation method is as follows~\cite{saez2001using}. The Saez tax utilizes income distribution $f(z)$ and cumulative distribution $F(z)$ to get the tax rates. The marginal tax rates denoted as $\tau(z)$, are expressed as a function of pretax income $z$, incorporating elements such as the income-dependent social welfare weight $G(z)$ and the local Pareto parameter $\alpha(z)$.
\begin{equation*}
    \tau(z)=\frac{1-G(z)}{1-G(z)+\alpha(z)e(z)}
\end{equation*}
To further elaborate, the marginal average income at a given income level $z$, normalized by the fraction of incomes above $z$, is denoted as $\alpha(z)$.
\begin{equation*}
    \alpha(z)=\frac{zf(z)}{1-F(z)}
\end{equation*}
The reverse cumulative Pareto weight over incomes above $z$ is represented by $G(z) $.
\begin{equation*}
    G(z)=\frac{1}{1-F(z)} \int_{z^{\prime}=z}^{\infty} p\left(z^{\prime}\right) g\left(z^{\prime}\right) \mathrm{d} z^{\prime}
\end{equation*}
From the above calculation formula, we can calculate $G(z) $ and $\alpha(z)$ by income distribution. We obtain the data of income and marginal tax rate through the interaction between the agent and environment and store them in the buffer. It is worth noting that the amount of buffer is fixed. 

To simplify the environment, we discretize the continuous income distribution, by dividing income into several brackets and calculating a marginal tax rate $\tau(z)$ for each income range.
Within each tax bracket, we determine the tax rate for that bracket by averaging the income ranges in that bracket. In other words, income levels falling within the income range are calculated as the average of that range.
In particular, when calculating the top bracket rate, it is not convenient to calculate the average because its upper limit is infinite. So here $G(z)$ represents the total social welfare weight of incomes in the top bracket, when calculating $\alpha(z)$, we take the average income of the top income bracket as the average of the interval.

Elasticity $e(z)$ shows the sensitivity of the agent's income z to changes in tax rates.
Estimating elasticity is very difficult in the process of calculating tax rates, here we estimate the elasticity $e(z)$ using a regression method through income and marginal tax rates under varying fixed flat-tax systems, which produces an estimate equal to approximately 1.
\begin{equation*}
    e(z)=\frac{1-\tau(z)}{z}\frac{dz}{d(1-\tau(z))}
\end{equation*}
\begin{equation*}
   \log (Z)=\hat{e} \cdot \log (1-\tau)+\log \left(\widehat{Z}^0\right)
\end{equation*}
where $Z=\sum_iz_i$ when tax rates is $\tau$.

\section{Game Theory Metrics}\label{metrics}
We will utilize the following metrics related to game theory to evaluate the effectiveness of the leader and follower policies: (1) The leader's payoff, which indicates the performance of the leader's policy in optimizing the leader's objective; (2) Exploitability, which measures the deviation of the agent's policy from the best response; (3) Social welfare, which assesses the deviation of the current state from the social optimum.
\paragraph{Leader's Payoff} We define the leader's payoff using the long-term expected rewards of the leader's policy \(\pi^l\) over discrete timesteps, as detailed in Equation~\ref{leader_utility}.

\paragraph{Exploitability} 
Exploitability is a critical metric in evaluating the convergence of policies and quantifying the divergence from the best response strategy in game theory. For a follower, exploitability $\mathcal{E}^f(\pi^f; \pi^l)$ is defined as the difference in payoffs between the follower's actual policy $\pi^f$ and its optimal response ${\pi^f}^*$, given the leader’s policy $\pi^l$. Formally, it is represented as:
\begin{equation*}
    \mathcal{E}^f(\pi^f; \pi^l) = J^f(\pi^l, {\pi^f}^*, L) - J^f(\pi^l, \pi^f, L),
\end{equation*}
where $J^f$ denotes the cumulative reward for the follower, defined in Section~\ref{fmfg}.

Similarly, the leader's exploitability $\mathcal{E}^l(\pi^l; \pi^f)$ measures the payoff difference between the leader’s policy $\pi^l$ and its best response ${\pi^l}^*$, given the followers' response policy $\pi^f$ and state-action distribution $L$. This is given by:
\begin{equation*}
    \mathcal{E}^l(\pi^l; \pi^f) = J^l(\pi^{l*}, \pi^f, L) - J^l(\pi^l, \pi^f, L),
\end{equation*}
with $J^l$ representing the cumulative reward for the leader, and $L = \{L_t\}_{t=0}^T$ detailing the state-action distribution for followers over time (see Section~\ref{fmfg}).

The overall exploitability, which measures the discrepancy from Nash equilibrium for both the leader and the followers, is defined as:
\begin{equation*}
    \mathcal{E}(\pi^l, \pi^f) =  \mathcal{E}^f(\pi^f; \pi^l) +  \mathcal{E}^l(\pi^l; \pi^{f}),
\end{equation*}
A near-zero value of $\mathcal{E}(\pi^l, \pi^f)$ indicates that the policies of both the leader and the followers are approaching their respective optimal strategies ${\pi^l}^*$ and ${\pi^f}^*$, signifying an equilibrium state.

\paragraph{Social Optimum and Social Welfare} In economic theory, the \textit{Social Optimum} describes a state in which the allocation of resources achieves maximum efficiency, as measured by social welfare~\cite{arrow1954existence,malul2009gap}. Given the leader's policy $\pi^l$ and the representative follower's policy $\pi^f$ among large-scale followers, social welfare $\mathcal{SW}(\pi^l, \pi^f)$ is approximately calculated as the sum of the utility functions defined in Section~\ref{modeling} of the $N$ followers:
\begin{equation*}
\begin{aligned}
    \mathcal{SW}(\pi^l, \pi^f) &= \sum^N_{i=1} J^{fi} \left(\pi^l, \pi^f, L\right)\\
    &= \mathbb{E}_{s_0^f \sim \mu_0^f, \mathbf{s}_{t+1} \sim P} \left[\sum_{t=0}^T \sum^N_{i=1} r^{fi}(\mathbf{s}_t,a^l_t, a^{fi}_t, L_t) \right]
\end{aligned}
\end{equation*}

\section{Additional Results}\label{appendix_results}
\subsection{Compute Resources}\label{resource} 
All experiments are run on 2 workstations: A 64-bit server with dual AMD EPYC 7742 64-Core Processors @2.25 GHz, 256 cores, 512 threads, 503GB RAM, and 2 NVIDIA A100-PCIE-40GB GPU. A 64-bit workstation with Intel Core i9-10920X CPU @ 3.50GHz, 24 cores, 48 threads, 125 GB RAM, and 2 NVIDIA RTX2080 Ti GPUs. The following Table~\ref{tab:experiment_results} shows the approximate training times for several algorithms.
\begin{table*}[h]
\centering
\begin{tabular}{ccccccc}
\toprule
\textbf{Algorithm} & \multicolumn{2}{c}{\textbf{Training Time (hours)}} & \multicolumn{2}{c}{\textbf{Utility (years)}} & \multicolumn{2}{c}{\textbf{Utility per training time}} \\
\cmidrule(r){2-3} \cmidrule(r){4-5} \cmidrule(r){6-7} 
 & \textbf{N=100} & \textbf{N=1000} & \textbf{N=100} & \textbf{N=1000} & \textbf{N=100} & \textbf{N=1000} \\
\midrule
DSMFG & 4 & 14 & 300 & 300 & \textbf{75.00} & \textbf{21.43} \\
DSMFG w/o MF & 3.5 & 16 & 1 & 1.02 & 0.29 & 0.06 \\
DSMFG w/o S & 4 & 9 & 75.75 & 1.5 & \underline{18.94} & 0.17\\
DSMFG w/o MF \& S & 2 & 6 & 1 & 1 & 0.50 & 0.17 \\
Free Market & 0.25 & 2 & 1 & 1 & 4.00 & 0.50 \\
Saez Tax & 4 & 23  & 300 & 100.58 & \textbf{75.00} & \underline{4.37} \\
AI Economist & 6.5 & N/A & 1 & N/A & 0.15 & N/A \\
\bottomrule
\end{tabular}
\caption{The average training times, utility (Years), and utility per training time for baselines in our experiments. The best values are highlighted in \textbf{bold}, and the second-best values are \underline{underlined}. Utility is measured using the "Years" metric, which represents the number of simulation steps achievable under a given policy. A higher number of simulation steps indicates better policy performance but also corresponds to increased computational complexity.}
\label{tab:experiment_results}
\end{table*}

\subsection{Further Experiments on the Necessity of SMFG} \label{smfg_nece}
\begin{figure*}[t]
    \centering
    \includegraphics[width=1\linewidth]{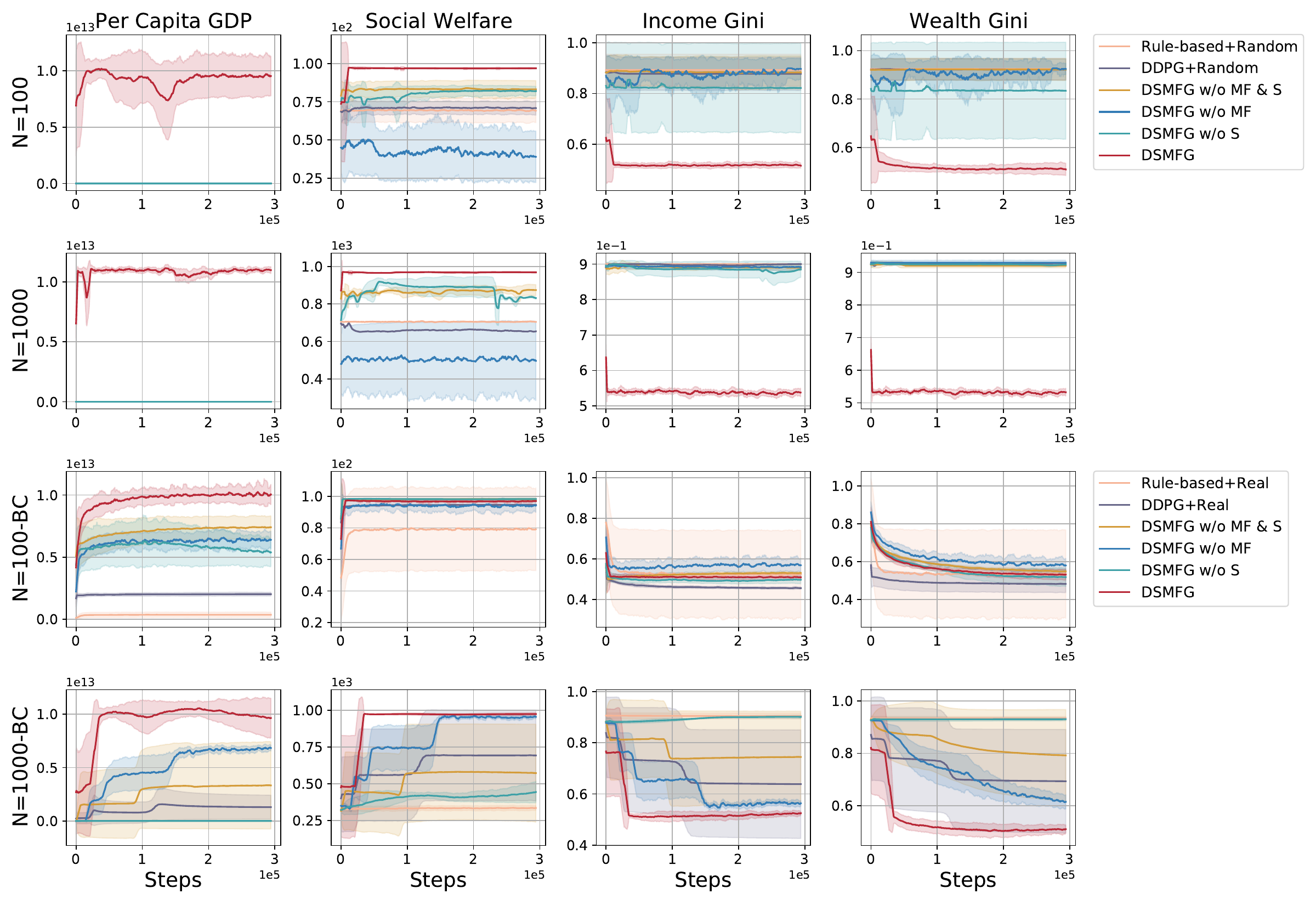}
    \caption{The training curves for 6 algorithms on 4 macroeconomic indicators, comparing settings without behavior cloning as pre-train (N=100 \& N=1000) and with behavior cloning (N=100-BC \& N=1000-BC).}
    \label{fig:n_100_training_full}
\end{figure*}

In this section, we present additional experimental results for validating the necessity of SMFG, including training curves Figure~\ref{fig:n_100_training_full} and Table~\ref{tab:policy_comparison_n100} and \ref{tab:policy_comparison_n1000}, as well as experiments incorporating the use of behavior cloning as a pre-training strategy for follower agents. We find that the DSMFG method without behavior cloning as pre-training still surpasses other baselines that utilize behavior cloning. More specifically, we compared DSMFG with 5 baselines across 4 different experimental setups: without behavior cloning as pre-training for follower agents at N=100 and N=1000 (marked as \textit{N=100 without BC} and \textit{N=1000 without BC}); with BC-based pre-training for follower agents at N=100 and N=1000 (\textit{N=100-BC}; \textit{N=1000-BC}).
Figure~\ref{fig:n_100_training_full} illustrates the training curves of 4 key macroeconomic indicators under these four settings. The solid line represents the average value of the metrics across the 5 random seeds, while the shaded area represents the standard deviation. Each row corresponds to one setting, and each column to a macroeconomic indicator, including per capita GDP, social welfare, income Gini, and wealth Gini. A rise in per capita GDP indicates economic growth, an increase in social welfare implies happier households and a lower Gini index indicates a fairer society. Each subplot's Y-axis represents the indicators' values, and the X-axis represents the training steps. Table~\ref{tab:policy_comparison_n100} and \ref{tab:policy_comparison_n1000} displays the test results of the 7 algorithms across 4 indicators, with each column corresponding to an experimental setting.

 Figure~\ref{fig:n_100_training_full} and Table~\ref{tab:policy_comparison_n100} and \ref{tab:policy_comparison_n1000} present two experimental findings: (1) Using BC as a pre-training method for the follower's policy enhances the algorithms' stability and performance. Comparing settings with and without BC (the first two rows), our method, DSMFG, shows similar convergence outcomes; however, the performance of other algorithms significantly improves across all four indicators with BC-based pre-training. Furthermore, the training curves of each algorithm are more stable. (2) The DSMFG method substantially outperforms other algorithms in solving DSMFGs, both in large-scale followers and without pre-training scenarios. In the setting of \textit{N=100-BC}, DSMFG achieved a significantly higher per capita GDP compared to other algorithms, while its social welfare and Gini index are similar to others, essentially reaching the upper limit. Besides, in \textit{N=100 without BC} and \textit{N=1000-BC}, DSMFG consistently obtains the most optimal solutions across all indicators.

\begin{table*}[t]
\centering
\caption{Performance of multiple policies on key macroeconomic indicators for $N=100$ households. The best values are highlighted in \textbf{bold}, and the second-best values are \underline{underlined}.}
\label{tab:policy_comparison_n100}
\begin{tabular}{ll l c c c c c}
\toprule
\textbf{Category}    & \textbf{Subcategory}  & \textbf{Policies}        & \textbf{Per Capita} & \textbf{Social} & \textbf{Income} & \textbf{Wealth} & \textbf{Years $\uparrow$} \\ 
 &  &         & \textbf{GDP $\uparrow$} & \textbf{Welfare $\uparrow$} & \textbf{Gini $\downarrow$} & \textbf{Gini $\downarrow$} &  \\ 
\midrule
\multirow{2}{*}{Static Policy} 
& Non-intervention        & Free Market & 1.37e+05    & 32.97      & 0.89      & 0.92      & 1.10\\ 
& Real-data   & US Federal Tax         & 4.88e+11    & 94.19      & 0.40      & 0.40      & 289.55         \\ 
 \midrule
& Rule-based  & Saez Tax & 2.34e+12 & 73.82  & \textbf{0.21} & \textbf{0.38} & \textbf{300.00}  \\  
\multirow{3}{*}{Dynamic Policy}
& Independent-based        & AI Economist& 1.26e+05    & 72.81      & 0.88      & 0.91      & 1.00\\ 
\multirow{3}{*}{without BC} 
& \multirow{4}{*}{Game-based} 
& Markov Game& 1.21e+05    & 83.09      & 0.88      & 0.92      & 1.00\\ 
&  & Stackelberg Game  & 1.23e+05    & 48.17      & 0.89      & 0.93      & 1.00\\ 
&  & Mean Field Game     & 8.66e+07    & 82.02      & 0.82      & 0.83      & 75.75          \\ 
&  & DSMFG (ours)       & \underline{9.59e+12}      & \underline{96.87}        & {0.52}        & {0.51}        & \textbf{300.00} \\ 
\midrule
\multirow{3}{*}{Dynamic Policy} 
& Independent-based        & AI Economist  & 2.03e+12    & 94.50      & \underline{0.46}        & \underline{0.48}        & \underline{299.85}         \\ 
\multirow{3}{*}{with BC}
& \multirow{4}{*}{Game-based} 
 & Markov Game& 7.41e+12    & 98.16      & 0.53      & 0.55      & \textbf{300.00} \\ 
&  & Stackelberg Game  & 6.38e+12    & 93.89      & 0.57      & 0.58      & 268.53         \\ 
&  & Mean Field Game     & 5.44e+12    & \textbf{98.21}        & 0.50      & 0.52      & \textbf{300.00} \\ 
&  & DSMFG (ours)      & \textbf{1.01e+13}      & 96.90      & 0.51      & 0.53      & \underline{299.89}         \\ 
\bottomrule
\end{tabular}%

\end{table*}

\begin{table*}[h]
\centering
\caption{Performance of multiple policies on key macroeconomic indicators for $N=1000$ households. The best values are highlighted in \textbf{bold}, and the second-best values are \underline{underlined}.}
\label{tab:policy_comparison_n1000}
\begin{tabular}{ll l c c c c c}
\toprule
\textbf{Category}    & \textbf{Subcategory}  & \textbf{Policies}        & \textbf{Per Capita} & \textbf{Social} & \textbf{Income} & \textbf{Wealth} & \textbf{Years $\uparrow$} \\ 
 &  &         & \textbf{GDP $\uparrow$} & \textbf{Welfare $\uparrow$} & \textbf{Gini $\downarrow$} & \textbf{Gini $\downarrow$} &  \\ 
\midrule
\multirow{2}{*}{Static Policy} 
& Non-intervention        & Free Market & 1.41e+05    & 334.79      & 0.90      & 0.93      & 1.00\\ 
& Real-data   & US Federal Tax         & 1.41e+05    & 351.17      & 0.89      & 0.93      & 1.00\\ 
 \midrule
 & Rule-based  & Saez Tax & 6.35e+11 & 498.88 & 0.68  & 0.73  & 100.58 \\ 
\multirow{3}{*}{Dynamic Policy} 
& Independent-based        & AI Economist& {N/A}       & {N/A}       & {N/A}     & {N/A}     & {N/A}          \\ 
\multirow{3}{*}{without BC} 
& \multirow{4}{*}{Game-based} 
 & Markov Game& 1.33e+05    & 874.53      & 0.89      & 0.92      & 1.00\\ 
&  & Stackelberg Game  & 1.48e+05    & 499.01      & 0.89      & 0.93      & 1.02\\ 
&  & Mean Field Game     & 1.31e+05    & 834.93      & 0.88      & 0.92      & 1.50          \\ 
&  & DSMFG (ours)       & \textbf{1.10e+13}      & \underline{968.94}        & \underline{0.54}        & \underline{0.53}        & \textbf{300.00} \\ 
\midrule
\multirow{3}{*}{Dynamic Policy} 
& Independent-based        & AI Economist   & {N/A}       & {N/A}       & {N/A}     & {N/A}     & {N/A}            \\ 
\multirow{3}{*}{with BC} 
& \multirow{4}{*}{Game-based} 
 & Markov Game& 2.79e+12    & 512.19      & 0.77      & 0.81      & 100.68         \\ 
&  & Stackelberg Game  & 6.82e+12    & 954.88      & 0.56      & 0.62      & \underline{278.50}         \\ 
&  & Mean Field Game     & 1.13e+05    & 440.00      & 0.90      & 0.93      & 1.00\\ 
&  & DSMFG (ours)       & \underline{9.68e+12}      & \textbf{975.15}        & \textbf{0.52}        & \textbf{0.51}        & \textbf{300.00} \\ 
\bottomrule
\end{tabular}%

\end{table*}

\subsection{Training Curves for Various Tax Policies}\label{training_tax}
\begin{figure*}
    \centering
    \includegraphics[width=1\linewidth]{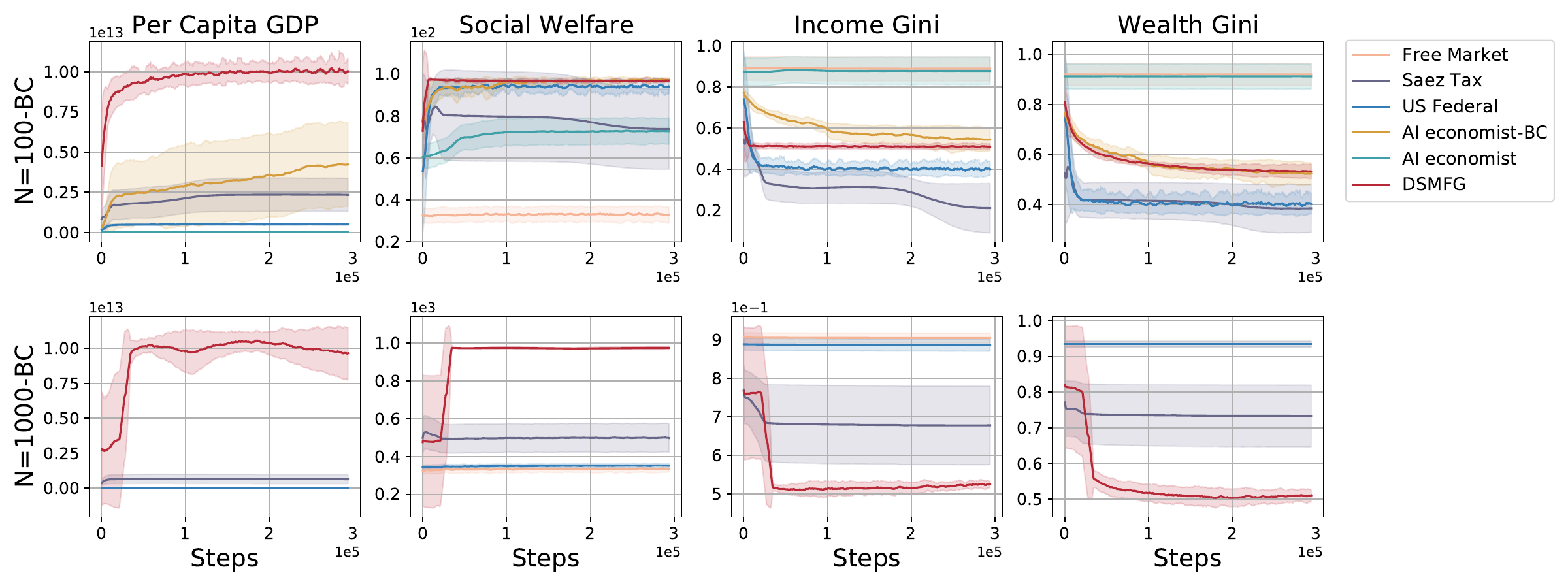}
    \caption{The training curves for 6 tax policies on 5 macroeconomic indicators (N=100 \& N=1000 with BC)}
    \label{fig:ai_policy}
\end{figure*}

We compare the performance of 6 policies across four economic indicators under two settings: with N=100 and N=1000 households. Figure~\ref{fig:ai_policy} displays the training curves and Table~\ref{tab:policy_comparison_n100} and \ref{tab:policy_comparison_n1000} shows the test results. Both Figure~\ref{fig:ai_policy} and Table~\ref{tab:policy_comparison_n100} and \ref{tab:policy_comparison_n1000} indicate that the DSMFG method significantly surpasses other policies in the task of optimizing GDP, and achieves the highest social welfare. When N=100, the Saez tax achieves the lowest income and wealth Gini coefficients, suggesting greater fairness. However, at N=1000, DSMFG performs optimally across all economic indicators, while the effectiveness of other policies noticeably diminishes as the number of households increases. The Saez tax also reduces the Gini index, but not as effectively or stably as the DSMFG.








\subsection{Efficiency-Equity Tradeoff of Policies}\label{tradeoff}
\begin{figure*}
    \centering
    \includegraphics[width=0.8\linewidth]{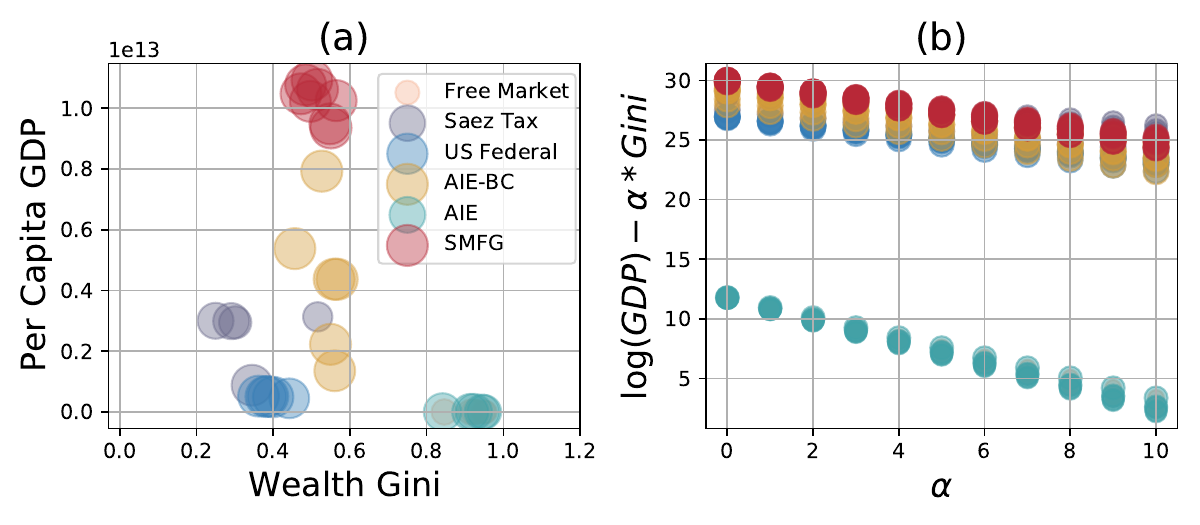}
    \caption{Comparative performance of various policies under multi-objective assessment (Efficiency-Equity).}
    \label{fig:tradeoff}
\end{figure*}
In economics, the Efficiency-Equity Tradeoff is a highly debated issue. We find that our DSMFG method is optimal in balancing efficiency-equity, except in cases of extreme concern for social fairness.
In our study, we depict the economic efficiency (Per capita GDP) on the Y-axis and equity (wealth Gini) on the X-axis of Figure~\ref{fig:tradeoff}(a) for various policies. Different policies are represented by circles of different colors, with their sizes proportional to social welfare. Different circles of the same color correspond to different seeds. Figure~\ref{fig:tradeoff} (a) shows that the wealth Gini indices for DSMFG and AI Economist-BC are similar, but DSMFG has a higher GDP, suggesting its superiority over AI Economist-BC. DSMFG significantly outperforms the free market policy and AI Economist due to its higher GDPs and lower wealth Ginis. However, comparing DSMFG with the Saez tax and the U.S. Federal tax policy in terms of both economic efficiency (GDP) and social equity (Gini) is challenging. Therefore, we introduce Figure~\ref{fig:tradeoff} (b) to demonstrate the performance of different policies under various weights in a multi-objective assessment.

In Figure~\ref{fig:tradeoff} (b), the Y-axis shows the weighted values of the multi-objective function $Y = log(\text{per capita GDP}) + \alpha (\text{wealth Gini})$, and the X-axis represents the weight of the wealth Gini index. For each weight $\alpha$, we compute the multi-objective weighted values for those policies, represented as circles of different colors. Due to the logarithmic treatment of GDP in (b), when $\alpha=10$, the overall objective focuses solely on social fairness; when $\alpha=0$, the overall objective is concerned only with efficiency.
Our findings in (b) reveal that only when $\alpha \geq 8$, which indicates a substantial emphasis on social equity, does the Saez tax outperform DSMFG. However, DSMFG consistently proves to be the most effective policy under a wide range of preference settings.


\subsection{Behavior Cloning Experiments}\label{bc_detail}

We conduct behavior cloning based on real data to simulate the behavior strategies of households in realistic scenarios, which are then used in Experiment~\ref{heter} to compare with DSMFG followers. We collect the statistical data from the 2022 Survey of Consumer Finances (SCF) ({https://www.federalreserve.gov/econres/scfindex.htm}) as the real data buffer $\mathcal{D}_{real}$.

Based on real data, we fetch a large number of followers' state-action pairs $\{s^f, a^f\}$ from a real-data buffer $\mathcal{D}_{real}$ for behavior cloning. For different settings of network structures, we have chosen two types of loss: when the neural network outputs a probability distribution of actions, we use the negative log-likelihood loss (NLL loss); when the neural network outputs action values, we employ the mean square error loss (MSE loss). Our goal is to find the optimal parameters $\theta$ as the follower's policy network $\pi_{\theta}$ initialization, thereby minimizing the loss to its lowest convergence.
\begin{equation*}
\begin{aligned}
   \min_{\theta}\mathcal{L}_{NLL} &= - \mathbb{E}_{s^f,a^f \sim \mathcal{D}}\log{\pi_{\theta}( a^f \mid s^f)}, \\
    \min_{\theta}\mathcal{L}_{MSE} &= \mathbb{E}_{s^f,a^f \sim \mathcal{D}} \left( a^f - a \right)^2 |_{a=\pi_{\theta}(s^f)}.
\end{aligned}
\end{equation*}

\begin{figure*}
    \centering
    \includegraphics[width=0.8\linewidth]{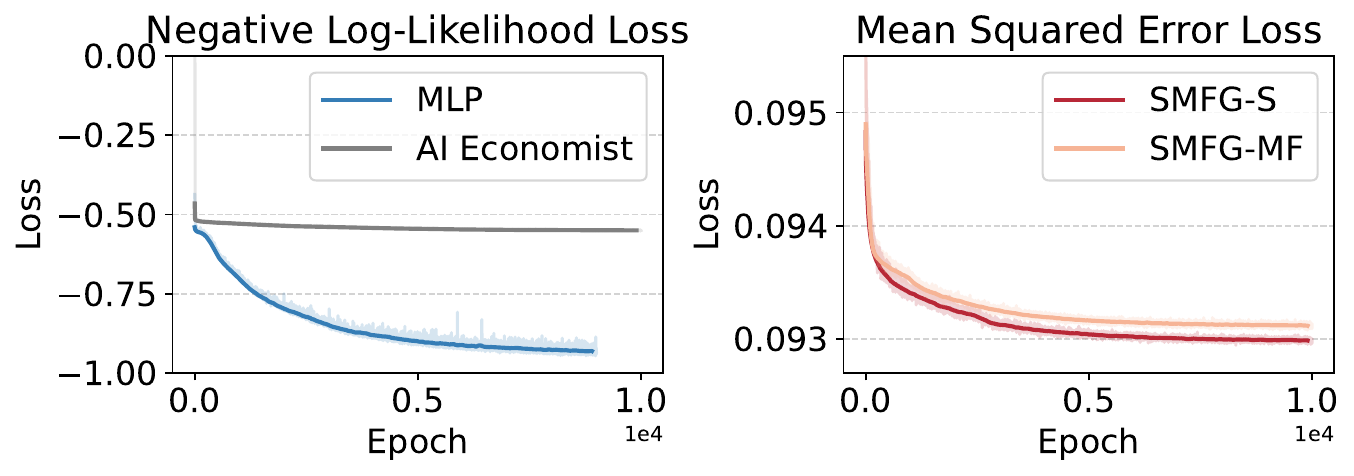}
    \caption{The behavior cloning loss for 4 networks in two loss types.}
    \label{bc-loss}
\end{figure*}

This experiment conducts behavior cloning on networks for four different household policies: Multilayer Perceptron (MLP), AI economist's network (MLP+LSTM+MLP), DSMFG w/o S, and DSMFG w/o MF network. The first two, as their network outputs, are probability distributions, use negative log-likelihood loss (Figure~\ref{bc-loss} left); the latter two's networks employ deterministic policies, hence they use mean square error loss against real data (Figure~\ref{bc-loss} right). The loss convergence curve of behavior cloning is shown in Figure~\ref{bc-loss}. It can be observed that the AI economist's network, due to its complexity, struggles to converge to near -1 like MLP. The losses corresponding to MFRL and DSMFG w/o MF can converge to below 0.1.

\section{TaxAI}
\subsection{Introduction of TaxAI}\label{intro_taxai}
TaxAI is a novel Multi-Agent Reinforcement Learning (MARL) environment designed to model dynamic interactions among governments, households, firms, and financial intermediaries. Built on the Bewley-Aiyagari economic model, TaxAI addresses critical limitations of existing economic simulators by offering enhanced scalability, realism, and benchmarking capabilities.  

\begin{itemize}
    \item Scalability: TaxAI simulates dynamic interactions involving up to 10,000 households, significantly surpassing the scale of prior simulators and enabling large-scale analysis. 
    \item Realism: Calibrated with real-world data from the 2022 Survey of Consumer Finances (SCF), TaxAI ensures its simulations reflect realistic economic scenarios, improving the relevance of its outcomes for policymaking.  
    \item Benchmarking: TaxAI evaluates 7 MARL algorithms against 2 traditional economic approaches (e.g., genetic algorithms, dynamic programming), demonstrating the superiority of MARL in addressing dynamic, partially observable economic environments.  \item Policy Optimization: TaxAI leverages MARL's adaptive learning capabilities to model complex government-household interactions and discover optimal tax policies that promote growth and equity.  
\end{itemize}
With its ability to integrate scalability, real-world calibration, and MARL-based adaptive optimization, TaxAI sets a new benchmark for realistic and effective economic simulators, providing actionable insights for policy design and implementation. Therefore, TaxAI is highly suitable as the experimental environment for this paper, particularly due to its scalability and realism.

\subsection{Economic model details}
Economic activities among households aggregate into labor markets, capital markets, goods markets, etc. In the labor market, households are the providers of labor, with the aggregate supply $ S(W_t) = \sum_i^N e_t^i h_t^i $, and firms are the demanders of labor, with the aggregate demand $ D(W_t) = \mathcal{L}_t $. When supply equals demand in the labor market, there exists an equilibrium price $ W_t^* $ that satisfies:
\begin{equation*}
S(W^*_t) = D(W^*_t), \mathcal{L}_t = \sum^N_i e_t^i h_t^i.
\end{equation*}

In the capital market, financial intermediaries play a crucial role, lending the total deposits of households $A_{t+1} = \sum_i^N a_{t+1}$ to firms as production capital $K_{t+1}$, and purchasing government bonds $B_{t+1}$ at the interest rate $r_t$. The capital market clears when supply equals demand:
\begin{equation*}
   K_{t+1}+B_{t+1}-A_{t+1}=\left(r_t+1\right) (K_t +  B_t-A_t )
\end{equation*}
In the goods market, firms produce and supply goods, while all households, the government, and physical capital investments $X_t$ demand them. The goods market clears when:
\begin{equation*}
    Y_t =C_t + G_t + X_t
\end{equation*}
where $C_t = \sum_i^N c^i_t$ represents the total consumption of households, and $G_t$ is government spending. The supply, demand, and price represent the states of the market. 

\subsection{Economic Shocks}\label{economic_shock}
In Experiment~\ref{experiment}, we simulate economic shocks analogous to a financial crisis: at the 100-th step, the wealth of all households is reduced by 50\%. In our economic model, this scenario is mathematically represented as follows: for each household member, the wealth \( a_t^i \) at time \( t \) is updated according to the rule
\[
a_t^i = 0.5  a_{t-1}^i, \quad \forall i \in \{1, \ldots, N\}
\]
where \( N \) denotes the total number of household members.

\section{Hyperparameters}\label{parameters}

\begin{table}[ht]
\label{table: alg_hyperparameter}
\begin{center}
\begin{tabular}{l r}
\toprule
Hyperparameter & {Value} \\  
\midrule
{Discount factor $\gamma$}  & 0.975 \\
{Replay buffer size}   & 1e6   \\
{Num of epochs }  & 1000  \\
{Epoch length}  &  300  \\
{Batch size}  &  128  \\
{Adam epsilon} & 1e-5\\
{Update cycles} & 100 \\
{Evaluation epochs}  & 10  \\
{Hidden size} & 128 \\
{Tau} & 0.95 \\ 
{Critic initial learning rate} & 3e-4 \\
{Actor initial learning rate}  & 3e-4 \\
{Learning rate adjustment} & $0.95^{(\text{epoch} // 35)}$\\
\bottomrule
\end{tabular}
\end{center}
\caption{Hyperparameters of DSMFG methods and its variants.}
\end{table}

\begin{table}[ht]
\label{table: MADDPG}
\begin{center}
\begin{tabular}{l r}
\toprule
Hyperparameter & {Value} \\  
\midrule
{Noise rate }&0.01 \\
{Epsilon start} & 0.1\\
{Epsilon end} & 0.05\\
{Epsilon decay} & 1e-5\\
\bottomrule
\end{tabular}
\end{center}
\vskip -0.1in
\caption{Hyperparameters of DSMFG w/o MF algorithm different from DSMFG method.}
\end{table}

\begin{table}[ht]
\begin{center}
\begin{tabular}{l r}
\toprule
Hyperparameter & {Value} \\  
\midrule
  {Tau $\tau$}& 5e-3\\
  {Gamma $\gamma$}&0.95\\
  {Eps $\epsilon$}&1e-5\\
  {Clip}&0.1 \\
  {Vloss coef}&0.5 \\
  {Ent coef} &0.01 \\
  {Government's initial learning rate} & 3e-4 \\
  {Learning rate adjustment}& 0, epoch $<$ 10 \\
  {} & $0.97^{(\text{epoch} // 35)}$, epoch $\geq$ 10 \\
  {Households' initial learning rate} & 1e-6 \\
  {Learning rate adjustment}&  $0.97^{(\text{epoch} // 35)}$ \\
\bottomrule
\end{tabular}
\end{center}
\vskip -0.1in
\caption{Hyperparameters of AI Economist Algorithm different from DSMFG approach.}
\end{table}

\section*{Ethical Statement}\label{impact}

This research introduces a novel DSMFG method, designed to optimize macroeconomic policies by modeling complex interactions at the micro level. The potential impact of this work extends across several domains:
\paragraph{Academic Contributions} The framework and algorithm proposed represent significant advancements in AI for economics and AI for social impact field, potentially serving as foundational tools for future research in macroeconomic policy making. By addressing the Lucas critique through dynamic modeling of individual agents within a mean field game, this work encourages more accurate and robust economic predictions and policy evaluations.

\paragraph{Policy Making and Societal Impact}
By enabling the optimization of macroeconomic policies through real-time, dynamic responses of micro-agents, this model provides policymakers with a powerful tool for assessing the impact of different economic strategies, leading to more informed decisions that maximize social welfare and economic stability, particularly in response to economic shocks. The application of this model can have profound implications for wealth distribution and social equity, helping ensure that economic policies are beneficial to a broader section of the population, potentially reducing inequality and enhancing societal well-being.

\paragraph{Ethical Considerations}
While the model aims to improve economic outcomes, the manipulation of macroeconomic policies must be approached with caution to avoid unintended negative consequences such as increased inequality or destabilization of economic sectors. Further, the reliance on AI-based decisions necessitates continuous scrutiny to ensure that the model accurately represents all population segments.

\paragraph{Limitations and Risks}
The complexity of the models also introduces risks related to the oversimplification of real-world dynamics and potential biases in the simulation of economic responses. Continuous validation against empirical data and diverse economic scenarios is essential to ensure the reliability and ethical application of the proposed methods.

In summary, the proposed DSMFG framework and SMFRL algorithm hold the potential to significantly impact both academic research and practical policy making, offering a new perspective on dynamic economic modeling that prioritizes realistic, individual-level responses within large-scale economic systems.

\end{document}